\documentclass[12pt, a4paper]{article}
\usepackage[height=21.5cm,width=16.5cm,centering]{geometry}

\usepackage{amsfonts}
\usepackage{amsmath}
\usepackage{amssymb}
\usepackage{ascmac}
\usepackage{dcolumn}
\usepackage{bm,here}
\usepackage{subfig}
\usepackage{comment}
\usepackage{ifpdf}
\usepackage{slashed}
\usepackage{colortbl}
\usepackage{xcolor}
\usepackage[mathscr]{eucal}
\usepackage[sort&compress, numbers, merge]{natbib}
\usepackage[utopia]{mathdesign}

\ifpdf        
  \usepackage{graphicx}     %   usepackage without driver option
  \usepackage[bookmarksopen,colorlinks=true,linkcolor=darkblue,citecolor=darkred,urlcolor=darkred,linktocpage=false]{hyperref}
\else     % For (p)LaTeX + dvipdfmx
  \usepackage[dvipdfmx]{graphicx}     %   usepackage with driver option
  \usepackage[dvipdfmx,bookmarksopen,colorlinks=true,linkcolor=darkblue,citecolor=darkred,
  urlcolor=darkred,linktocpage=false]{hyperref}
\fi

\usepackage{multicol}
\definecolor{red}{rgb}{1,0,0}
\definecolor{blue}{rgb}{0,0,1}
\definecolor{orange}{rgb}{1,0.5,0}
\definecolor{ppink}{rgb}{0.921545,0.440586,0.687243}
\definecolor{bblue}{rgb}{0.400000,0.400000,1.000000}
\definecolor{darkblue}{rgb}{0.0, 0.0, 0.6}
\definecolor{darkred}{rgb}{0.6,0,0}

\begin{document}

%%%%%%%%%%%%%%%%%%%%%%%%%%%%%%%%%%%%%%%%%%%%%%%%%%
\begin{titlepage}

%%Please delete it for utopia
%\newgeometry{height=21.5cm,width=18cm}

\begin{center}

\hfill UT 16-20\\
\hfill KEK-TH 1898\\
\hfill IPMU 16-0062\\

\vskip .75in

{\Large \bf 
Gravitational Particle Production in Oscillating Background\\[1em]
and Its Cosmological Implications
}

\vskip .75in

{\large Yohei Ema$^a$, Ryusuke Jinno$^b$, Kyohei Mukaida$^c$ and Kazunori Nakayama$^{a,c}$}

\vskip 0.25in

\begin{tabular}{ll}
$^{a}$&\!\! {\it Department of Physics, Faculty of Science, }\\
& {\em The University of Tokyo,  Bunkyo-ku, Tokyo 133-0033, Japan}\\[.3em]
$^{b}$&\!\! {\it Theory Center, KEK, 1-1 Oho, Tsukuba, Ibaraki 305-0801, Japan}\\[.3em]
$^{c}$ &\!\! {\it Kavli IPMU (WPI), UTIAS,}\\
&{\em The University of Tokyo,  Kashiwa, Chiba 277-8583, Japan}
\end{tabular}

\end{center}
\vskip .5in

\begin{abstract}

We study production of light particles due to oscillation
of the Hubble parameter or the scale factor.
Any coherently oscillating scalar field, irrespective of its energy fraction in the universe,
imprints such an oscillating feature on them.
Not only the Einstein gravity but extended gravity models, such as models with non-minimal (derivative) coupling
to gravity and $f(R)$ gravity, lead to oscillation of the scale factor.
We present a convenient way to estimate the gravitational particle production rate in these circumstances.
Cosmological implications of gravitational particle production, such as dark matter/radiation and moduli problem, are discussed.
For example, if the theory is described solely by the standard model 
plus the Peccei-Quinn sector, the Starobinsky $R^2$ inflation 
may lead to observable amount of axion dark radiation.

\end{abstract}

%%please delete it for utopia
%\restoregeometry

\end{titlepage}

\tableofcontents
\thispagestyle{empty}
\renewcommand{\thepage}{\arabic{page}}
\renewcommand{\thefootnote}{$\natural$\arabic{footnote}}
\setcounter{footnote}{0}
%%%%%%%%%%%%%%%%%%%%%%%%%%%%%%%%%%%%%%%%%%%%%%%%%%

\newpage
\setcounter{page}{1}

%%%%%%%%%%%%%%%%%%%%%%%%%%%%%%%%%%%%%%%%%%%%%%%%%%
\section{Introduction}
\label{sec:}
\setcounter{equation}{0}
%%%%%%%%%%%%%%%%%%%%%%%%%%%%%%%%%%%%%%%%%%%%%%%%%%
A scalar field often plays an important role in the cosmology. 
Inflaton, which drives the accelerated expansion in the early universe,
is an obvious example of such a scalar field~\cite{Guth:1980zm,Starobinsky:1980te}.
Curvaton~\cite{Lyth:2001nq,Moroi:2001ct}, 
which may offer seeds of the present 
large scale structure of the universe, is another example.
A scalar field can also generate the baryon to photon ratio 
in the present universe~\cite{Affleck:1984fy,Dine:1995kz}.
These scalar fields typically have to transfer their energy to other components of the universe, 
\textit{e.g.}, radiation, during their oscillating regimes.
Hence, particle production by a coherently oscillating background is rather common.
One of the most prominent examples is (p)reheating after inflation, 
caused by violent oscillation of inflaton.
See Refs.~\cite{Dolgov:1989us,Traschen:1990sw,
Shtanov:1994ce,Kofman:1994rk,Kofman:1997yn} for instance.

In most cases studied so far, 
an explicit coupling between the oscillating scalar field $\phi$ and another light field $\chi$ is introduced
in the action.
However, the simplest situation is that $\phi$ and $\chi$ interact only through gravity:
there are no interactions between them if the cosmic expansion is shut off,
or they interact only through the evolution of the universe.\footnote{
	In our terminology, if Planck-suppressed operators involving $\phi$ and $\chi$ are introduced explicitly in the action,
	they are regarded as explicit couplings, which are not of our interests.
}
It is known that even in such a case, particle production occurs 
through the change of the cosmic evolution caused by $\phi$ field,
which is often called ``gravitational particle production''~\cite{Ford:1986sy,Peebles:1998qn}.
Recently, we have pointed out that (small) oscillation of the Hubble parameter or the scale factor
caused by inflaton oscillation generates particles which couple to gravity non-conformally~\cite{Ema:2015dka}.
Such gravitational production takes place even in the Einstein gravity 
and its effect becomes stronger for some extended gravity theories.
Although the production rate is suppressed by the Planck scale, still it can have impacts on cosmology.

In this paper, we extend our previous analysis to cope with more general cases 
where the oscillation of the scale factor is caused by
coherent oscillation of an either dominant or subdominant scalar field. 
First we consider the system with the Einstein gravity 
and a scalar field coupled minimally with gravity.
Then gravity sectors are extended.
As examples, we consider $f(\phi)R$, $f(R)$ and also $G^{\mu\nu}\partial_\mu\phi\partial_\nu\phi$ models
where $\phi$ is the scalar field, $R$ is the Ricci scalar and $G^{\mu\nu}$ is the Einstein tensor, respectively.
In these models, the oscillation of the scale factor is more prominent than that in the Einstein gravity,
and hence gravitational particle production becomes more efficient.
We also discuss cosmological implications of the gravitational particle production
such as the dark matter/radiation and the moduli problem in each case.

Before starting the analysis, let us clarify differences of our study from existing literature.
For example, in Ref.~\cite{Watanabe:2007tf}, particle production of 
the $f(\phi)R$ theory is considered with $\phi$ being inflaton.
The authors in Refs.~\cite{Vilenkin:1985md,Mijic:1986iv,Arbuzova:2011fu} 
discussed particle production of the $f(R)$ theory
for the case of $f(R) = R + cR^2$ where the second term is dominant.
Therefore our study has some overlaps with these literature.
However, the goal of our paper is to offer a systematic way to
estimate the particle production rate.
For that purpose, we pay particular attention to an oscillating feature
of the scale factor. 
In fact, we will see that 
we can treat a wider class of gravity models 
in a unified way from this viewpoint.
It also makes manifest how the background oscillation produces particles
coupled with gravity non-conformally.
In addition to the above point, 
we also discuss the case where the oscillating scalar field is subdominant.
A subdominant scalar field can have some cosmological implications 
especially in the extended gravity theories as we will see below.

In this paper, we calculate the particle production rate in the original defining frame
(\textit{i.e.}, the Jordan frame for the $f(\phi)R$ and $f(R)$ theories).
This is partly because we cannot go to the Einstein frame in some classes of extended gravity, 
such as the $G^{\mu\nu}\partial_\mu\phi\partial_\nu\phi$ theory.
In order to estimate the production rate including such a case, 
we would like to understand what is happening in the original frame
in our unified framework.
Roughly speaking, we are seeking for effects of Planck-suppressed interactions of the oscillating scalar field
on particle production. 
As expected, it is not so violent compared with the preheating in the usual context~\cite{Traschen:1990sw,
Shtanov:1994ce,Kofman:1994rk,Kofman:1997yn}, but still it can play important role on cosmology,
e.g., dark matter/dark radiation production.
The effect is prominent in some extended gravity models
and the gravitational coupling itself can be the main source of reheating.

The organization of this paper is as follows.
In Sec.~\ref{sec:Ein}, we consider the Einstein gravity,
and show that
a (small) oscillating part of the scale factor is induced 
by the coherently oscillating scalar field even in such a minimal case.
We show that this process can be understood as annihilation of the scalar field.
In Sec.~\ref{sec:fphi}, we consider the $f(\phi)R$ theory.
In this case, we show that the oscillating part of the scale factor 
linearly depends on the scalar field in general, and
hence the scalar field can decay into light particles gravitationally,
in contrast to the annihilation process in the previous section.
In Sec.~\ref{sec:fR}, we consider the $f(R)$ theory.
We find that the situation is rather similar to that 
of the $f(\phi)R$ theory in this case.
In Sec.~\ref{sec:G}, we consider the
$G^{\mu\nu}\partial_{\mu}\phi\partial_{\nu}\phi$ theory.
In this case, we limit ourselves to the case
where the scalar field is subdominant to avoid a gradient instability.
Sec.~\ref{sec:con} is devoted to the conclusions and discussion.

%%%%%%%%%%%%%%%%%%%%%%%%%%%%%%%%%%%%%%%%%%%%%%%%%%
\section{Einstein gravity}  \label{sec:Ein}
%%%%%%%%%%%%%%%%%%%%%%%%%%%%%%%%%%%%%%%%%%%%%%%%%%
In this section, we consider gravitational particle production in the Einstein gravity.
We will show that the scale factor has an oscillating feature caused by the coherent oscillation
of a scalar field $\phi$ even if it is subdominant.
If it dominates the universe, 
the amount of produced particles, whose dominant contribution comes from the onset of its oscillation,
becomes comparable to 
that produced by the change of the background geometry \cite{Ford:1986sy,Ema:2015dka},
as expected.
%%%%%%%%%%%%%%%%%%%%%%%%%%%%%%%%%%%%%%%%%%%%%%%%%%
\subsection{Background dynamics}
%%%%%%%%%%%%%%%%%%%%%%%%%%%%%%%%%%%%%%%%%%%%%%%%%%

Let us consider the action
\begin{align}
	S = \int d^4 x \sqrt{-g}\left( \frac{1}{2}M_P^2 R - \frac{1}{2}(\partial \phi)^2 - V(\phi) + \mathcal L_M \right).
\end{align}
where $g = \det\, (g_{\mu\nu})$ is the determinant of the metric,
$M_P$ is the reduced Planck scale and $R$ is the Ricci scalar.
Here and hereafter, 
we adopt the $(-+++)$ convention for the metric $g_{\mu\nu}$.
The scalar field $\phi$, which is of our main interest,
oscillates coherently and imprints an oscillatory feature in the scale factor.
$\mathcal L_M$ denotes the Lagrangian for matter other than the scalar $\phi$.
We assume that $\mathcal L_M$ does not depend on $\phi$.

The background equation of motion of $\phi$ is given by
\begin{equation}
	\ddot \phi + 3H\dot\phi + V'= 0,
\end{equation}
where $H$ is the Hubble parameter
and the prime denotes the derivation with respect to $\phi$.
This is also rewritten as
\begin{equation}
	\dot \rho_\phi + 3H \left(\rho_\phi + p_\phi\right) = 0,
\end{equation}
where $\rho_\phi \equiv \dot\phi^2/2 + V$ and $p_\phi \equiv \dot\phi^2/2 - V$.
The Einstein equation reads
\begin{align}
	3H^2 &= \frac{\rho_\phi + \rho_M}{M_P^2},  \label{Ein_Fried}\\
	3H^2+2\dot H &= -\frac{p_\phi + p_M}{M_P^2},  \label{Ein_Raychaudhuri}
\end{align}
where
\begin{align}
	\rho_M =  g_{00}\mathcal L_M - 2\frac{\delta \mathcal L_M}{\delta g^{00}},~~~p_M = \mathcal L_M.
\end{align}
Note that $\delta \mathcal{L}_M/\delta g^{ij} = 0$ for the background part.
By using these equations, we obtain
\begin{equation}
	\dot \rho_M + 3H(\rho_M + p_M)=0. \label{drhoM}
\end{equation}
Hereafter we assume that the matter part satisfies the equation of state $p_M = w\rho_M$.
This shows that $\rho_M$ exactly scales as $\rho_M \propto a^{-3(1+w)}$.

The cosmological setup we are considering is as follows.
After inflation, inflaton decays and the universe is dominated by the ``matter''\footnote{
	Here and in what follows ``matter'' does not always mean non-relativistic fluids.
}
which is characterized by the energy density $\rho_M$ and the equation of state $w$.
We leave it as a free parameter for a while,
although hot thermal plasma with $w = 1/3$ is typically produced by the inflaton decay.
The scalar field $\phi$ begins to oscillate around the time $H=m_\phi$ in a background dominated by $\rho_M$,
with 
$m_\phi^2\equiv \left|(\partial V/\partial\Phi) / \Phi\right|$ being the effective mass squared of $\phi$
and $\Phi$ being the amplitude of $\phi$ oscillation.
In the following, we consider the deeply oscillating regime $m_\phi \gg H$.
We do not necessarily assume that $\phi$ dominates the universe in the following discussion.
Even if $\phi$ is subdominant, it induces a small oscillating feature in the Hubble parameter or the scale factor,
and leads to particle production as we will see below.

Henceforth,
we will extract an oscillating part of the scale factor which is important 
for the gravitational particle production.
Especially, we will express it in terms of $\phi$ explicitly.
To do so, we divide quantities into 
the oscillation-averaged part, 
which only evolves due to the Hubble expansion,
and rapidly oscillating part with frequency of order $\sim m_\phi$:
\begin{align}
	H &= \langle H\rangle + \delta H,\\
	a &= \langle a\rangle + \delta a,\\
	\rho_{\phi} &= \langle \rho_{\phi}\rangle + \delta \rho_\phi,\\
	\rho_{M} &= \langle \rho_{M}\rangle + \delta \rho_M.
\end{align}
Here the bracket $\langle ...\rangle$ denotes the oscillation average, and quantities with $\delta$
denote the oscillating part.
We treat the oscillating parts as perturbations, and keep only terms up to first order in them.
This treatment is justified in the deeply oscillating regime.

We first note that 
$\rho_M$ exactly scales as $\rho_M \propto a^{-3(1+w)}$, and also 
$\dot\rho_\phi \sim \mathcal O(H \rho_\phi)$.
Thus we can use the Virial theorem for $\phi$ in the limit $m_\phi \gg H$ and take oscillation average to obtain
\begin{equation}
	\dot{\langle \rho_{\phi}\rangle} + \frac{6n}{n+2} \langle H\rangle \langle \rho_{\phi}\rangle = 0,
\end{equation}
where we have assumed 
$V \sim \phi^n$ dominates in the potential.\footnote{
	If $\phi$ oscillates around the finite VEV $\phi_0$, $\phi$ should be interpreted as
	its deviation from the potential minimum.
}
This implies that $\rho_\phi$ scales as $\langle \rho_\phi\rangle \propto a^{-6n/(n+2)}$.
In order to extract the oscillating part, the following equation is useful:
\begin{align}
	\dot{H} = -\sum_{i = \phi, M} \frac{\rho_i + p_i}{2M_P^2}.
\end{align}
From this expression, we obtain the oscillating part of the Hubble parameter as
\begin{align}
	\dot{\delta H}
	\simeq -\frac{1}{2M_P^2}
	\left( \dot{\phi}^2 - \left\langle \dot{\phi}^2\right\rangle + (1+w)\delta \rho_M\right).
	\label{eq:ddH0}
\end{align}
Let us make an order-of-magnitude estimation to understand its approximated behavior.
We have $\delta\rho_M / \rho_M \sim \delta a/a \sim \mathcal O(\delta H / m_\phi)$
since the relation $\rho_M \propto a^{-3(1+w)}$ holds exactly.
Therefore, we obtain $\delta \rho_M /M_P^2 \lesssim \mathcal O(H^2 \delta H/m_\phi) 
\ll m_\phi \delta H$.
Thus, the last term in the RHS of Eq.~\eqref{eq:ddH0} can be neglected,
and hence we find
\begin{align}
	\dot{\delta H} \simeq - \frac{1}{2M_P^2}
	\left( \dot{\phi}^2 - \left\langle \dot{\phi}^2\right\rangle \right).
	\label{eq:dd}
\end{align}
It can be expressed as
\begin{align}
	\dot{\delta H} + \frac{6n}{n+2}\langle H\rangle \delta H \simeq -\frac{1}{n+2}\frac{1}{M_P^2}\left(\frac{d}{dt} + 3H\right)(\phi\dot\phi),
	\label{ddH20}
\end{align}
where we have used the oscillating part of Eq.~\eqref{Ein_Fried}.
A similar equation was derived in Ref.~\cite{Ema:2015dka} 
in the case where $\phi$ dominates the universe.
In contrast, here, we have not necessarily assumed $\phi$-domination.
In Eq.~\eqref{ddH20}, only the relevant terms are the first terms of LHS and RHS.
This is because $\delta H$ and $\phi\dot\phi$ are oscillating function with frequency $\sim m_\phi$,
and hence the second terms of LHS and RHS are suppressed by $\mathcal O (H/m_\phi)$.
Thus we arrive at
\begin{align}
	\delta H \simeq -\frac{1}{n+2}\frac{\phi\dot\phi}{M_P^2}.
	\label{eq:delH}
\end{align}
By integrating this, we obtain
\begin{equation}
	\frac{a(t)}{\langle a(t) \rangle} \simeq 1 - \frac{1}{2(n+2)} \frac{\phi^2 - \langle\phi^2\rangle}{M_P^2}.  \label{a0}
\end{equation}
This equation explicitly relates the oscillating part of the scale factor to the (subdominant) oscillating scalar field.
Note that $\phi^2$ appears regardless of the exponent $n$ of the potential.

%%%%%%%%%%%%%%%%%%%%%%%%%%%%%%%%%%%%%%%%%%%%%%%%%%
\subsection{Particle production rate}
%%%%%%%%%%%%%%%%%%%%%%%%%%%%%%%%%%%%%%%%%%%%%%%%%%
In the previous subsection, we obtain
\begin{align}
	a(t) \simeq \langle a (t) \rangle
	\left(1 - \frac{1}{2(n+2)} \frac{\phi^2 - \langle\phi^2\rangle}{M_P^2}\right).
\end{align}
Now let us estimate the particle production rate 
due to the oscillating part of the scale factor $a$.
Intuitively, in the present case,
such a particle production may be understood as pair annihilation of $\phi$,
since the oscillating part of the scale factor depends quadratically on $\phi$.
Thus we call it as ``gravitational annihilation''~\cite{Ema:2015dka}.\footnote{
	In Ref.~\cite{Watanabe:2007tf}, the gravitational annihilation due to $f(\phi)R$ coupling was discussed.
	There it was claimed that this effect does not exist in the Einstein gravity limit.
	This is not true, however, as shown here and also in 
	Ref.~\cite{Ema:2015dka}: the gravitational annihilation takes place
	even in the Einstein gravity.
	Recently Ref.~\cite{Garny:2015sjg,Tang:2016vch} considered dark matter production 
	by the gravitational annihilation of particles in thermal bath.
}
On the other hand, we will see that 
the oscillating part of the scale factor depends linearly on the scalar field
for extended gravity theories such as $f(\phi)R$ and $f(R)$ models.
In contrast to the present case, we can view it as decay of the scalar field,
and hence we will use the word ``gravitational decay'' in such cases.

Below we consider particle production of a minimally coupled scalar and the graviton.
The production of fermions and vector bosons is suppressed by their masses and couplings
because they are classically Weyl-invariant and do not 
feel the oscillation of the scale factor in the massless limit.

%%%%%%%%%%%%%%%%%%%%%%%%%%%%%%%%%%%%%%%%%%%%%%%%%%
\subsubsection{Scalar}
%%%%%%%%%%%%%%%%%%%%%%%%%%%%%%%%%%%%%%%%%%%%%%%%%%
First we consider a scalar field $\chi$ which minimally couples with gravity
\begin{align}
	S = \int d^4 x \sqrt{-g} \left( -\frac{1}{2}(\partial \chi)^2 - \frac{1}{2}m_\chi^2 \chi^2 \right),  \label{S_chi}
\end{align}
with $m_\chi \ll m_\phi$. In the standard model (SM), only the Higgs boson would be a minimally coupled scalar.
Here we do not limit ourselves to the case where $\chi$ is the Higgs boson, but consider general scalar fields.
By using the master formula~\eqref{nchi_grav_phi} derived in App.~\ref{sec:app},
the number density of $\chi$ particles produced during one Hubble time after $\phi$ begins to oscillate is given by\footnote{
	If $\phi$ dominates the universe, we have $n_\chi(t) \sim ({\rm const})\times H^3$ as found in \cite{Ford:1986sy,Ema:2015dka}.
}
\begin{align}
	n_\chi(t) \simeq \frac{C}{32\pi H}\left(\frac{1}{n+2}\right)^2\left(\frac{m_\phi^2\Phi^2}{M_P^2}\right)^2, 
	\label{nchi_scalar}
\end{align}
where $\Phi$ denotes the oscillation amplitude of $\phi$.
This result may be translated into the effective annihilation 
process of $\phi$ particles with a rate of
\begin{align}
	\Gamma_{\phi\phi\to\chi\chi} \equiv n_\phi\langle\sigma v\rangle_{\phi\phi\to\chi\chi} 
	\simeq \frac{C}{16\pi}\left(\frac{1}{n+2}\right)^2 \frac{\Phi^2}{M_P^2}\frac{m_\phi^3}{M_P^2}.
\end{align}
Taking into account the Hubble expansion, one can easily see
that the largest contribution comes from the very beginning of the $\phi$-oscillation
at $H=m_\phi$ unless $w$ is unlikely large and/or $n$ is so small.
Note that even if $\chi$ obtains a Hubble induced mass term, 
this production mechanism becomes effective soon after the $\phi$ oscillation.

%%%%%%%%%%%%%%%%%%%%%%%%%%%%%%%%%%%%%%%%%%%%%%%%%%
\subsubsection{Graviton}  \label{sec:grav_E}
%%%%%%%%%%%%%%%%%%%%%%%%%%%%%%%%%%%%%%%%%%%%%%%%%%

Next we apply our formalism to the graviton production.
The graviton action is given by
\begin{align}
	S = \int d\tau d^3x\, a^2(t)\frac{M_P^2}{8}
	\left[ \left(\frac{\partial h_{ij}}{\partial\tau}\right)^2 - (\partial_k h_{ij})^2 \right],
\end{align}
where $\tau$ is the conformal time, and $h_{ij}$ is the metric perturbation
satisfying the transverse and traceless conditions $h_{ii}=\partial_i h_{ij}=0$.
The indices $i,j$ and $k$ run the space coordinates.
Hence the production rate is similar to the minimal scalar,
except for the factor 2 corresponding to the two polarization states of the graviton:
\begin{equation}
	n_h(t) \simeq \frac{C}{16\pi H}\left(\frac{1}{n+2}\right)^2\left(\frac{m_\phi^2\Phi^2}{M_P^2}\right)^2. \label{nh_E}
\end{equation}
%%

%%%%%%%%%%%%%%%%%%%%%%%%%%%%%%%%%%%%%%%%%%%%%%%%%%
\subsection{Cosmological implications} \label{sec:cos_grav_E}
%%%%%%%%%%%%%%%%%%%%%%%%%%%%%%%%%%%%%%%%%%%%%%%%%%

The gravitational annihilation of a subdominant scalar field $\phi$ 
yields the abundance given in (\ref{nchi_scalar}), 
but the gravitational annihilation of inflaton also gives a significant contribution~\cite{Ford:1986sy,Ema:2015dka}.
The ratio of $\chi$ abundance produced by 
a subdominant scalar field $\phi$ to that produced by the inflaton is estimated as
\begin{align}
	\frac{n_\chi^{(\phi)}} {n_\chi^{\rm (inf)}}\simeq \epsilon\frac{m_\phi}{H_{\rm inf}}\left(\frac{\phi_i}{M_P}\right)^4,
\end{align}
where
\begin{align}
	\epsilon = {\rm min}\left[1,~ \sqrt{m_\phi/\Gamma_{\rm inf}}\right],
\end{align}
with $H_{\rm inf}$ and $\Gamma_{\rm inf}$ being the Hubble scale at the end of inflation and the inflaton decay rate, respectively.
Here we have assumed that the inflaton oscillation behaves 
as non-relativistic matter and the inflaton decays into radiation.
Also, $\phi$ is assumed to be sub-dominant at the onset of its oscillation.
The dominant contribution comes from the gravitational annihilation of the inflaton
since $m_\phi < H_{\rm inf}$ in order for $\phi$ to begin coherent oscillation after inflation.
In this case, cosmological implications were studied in \cite{Ema:2015dka},
and we briefly discuss them here.
The energy density-to-entropy ratio of $\chi$ with a sizable mass term
is estimated to be
\begin{align}
	\frac{\rho_\chi^{\rm (inf)}}{s} \simeq \frac{1}{\Delta} \frac{9C}{512\pi}\frac{m_\chi T_{\rm R}H_{\rm inf}}{M_P^2}
	\simeq 1\times 10^{-9}\,{\rm GeV}\frac{C}{\Delta}\left( \frac{m_\chi}{10^6\,{\rm GeV}} \right) 
	 \left( \frac{T_{\rm R}}{10^{10}\,{\rm GeV}} \right) 
	 \left( \frac{H_{\rm inf}}{10^{14}\,{\rm GeV}} \right),
\end{align}
where $s$ is the entropy density, $T_{\rm R}$ is the reheating temperature
and $\Delta$ denotes the dilution factor due to the late decay of $\phi$, which is given by\footnote{
	Here we have introduced some interactions that induce a complete decay of $\phi$,
	in order for the $\phi$ oscillation not to dominate the universe.
	For simplicity, we assume that this interaction does not involve $\chi$.
}
\begin{align}
	\Delta = {\rm max}\left[1,~ \sqrt{H_{\rm dom}/\Gamma_\phi}\right].
\end{align}
Here $H_{\rm dom}$ is the Hubble parameter when $\phi$ would dominate the universe:
$H_{\rm dom} = \Gamma_{\rm inf}(\phi_i^2/6M_P^2)^2$ for $m_\phi > \Gamma_{\rm inf}$ and
$H_{\rm dom} = m_\phi(\phi_i^2/6M_P^2)^2$ for $m_\phi < \Gamma_{\rm inf}$
when the exponent of the potential of $\phi$ is $n = 2$.

Suppose that $\chi$ is a massive non-interacting stable particle.
Then, its abundance should be smaller than $\rho_\chi/s \lesssim 4\times 10^{-10}\,{\rm GeV}$
to avoid the dark matter (DM) overproduction.\footnote{
	Depending on $m_\chi$ and $m_\phi$, the free-streaming length of $\chi$ can be so long
	that it fails to be a cold DM.
	In such a case, its abundance must be well below the observed DM abundance.
	In order for $\chi$ to be cold, it should become non-relativistic before the cosmic 
	temperature drops down to $\sim 1$\,keV. 
}
Next, suppose that $\chi$ is a moduli that has only Planck-suppressed interactions with SM fields.
It is severely constrained from cosmology due to its longevity.
If its mass is about $\mathcal O(1) $\,TeV, big-bang nucleosynthesis (BBN) gives a stringent bound on the $\chi$ abundance,
$\rho_\chi/s \lesssim 10^{-14}\,{\rm GeV}$ \cite{Kawasaki:2004qu}.
Thus we roughly have
\begin{align}
	\frac{\rho_\chi}{s} \lesssim  10^{-14} - 4\times 10^{-10} \,{\rm GeV},  \label{constraint}
\end{align}
depending on the mass, lifetime, decay modes etc. Various cosmological constraints on massive particles in broad parameter space are found in Ref.~\cite{Kawasaki:2007mk}.
There is no such constraint if $\chi$ decays well before BBN begins.
Finally, if $\chi$ is (nearly) massless, like an axion-like particle, it contributes to dark radiation.
In this case, however, the $\chi$ abundance as well as the gravitational wave abundance is so small that it does not affect observations.

Note also that $\chi$ can have either (dominantly) adiabatic or isocurvature fluctuation depending on 
whether $\phi$ is massive or not during inflation.
If $\phi$ remains light during inflation ($m_\phi \lesssim H_{\rm inf}$), it obtains long-wavelength quantum fluctuations and contributes to the curvature perturbation as
\begin{equation}
	\zeta_\phi \simeq \frac{R_\phi}{3} \frac{H_{\rm inf}}{\pi\phi_i},
\end{equation}
where $R_\phi$ is the fraction of $\phi$ energy density at its decay,
and it can act as the curvaton~\cite{Lyth:2001nq,Moroi:2001ct}. 
If $\phi$ is the dominant source of the curvature perturbation, 
$\chi$ produced by the inflaton oscillation has totally 
anti-correlated isocurvature perturbation and it cannot be the dominant component of DM~\cite{Lyth:2002my}. 
If the curvature perturbation is dominantly sourced by the inflaton, 
there is no significant constraint from the isocurvature perturbation.
Also, if $\phi$ is heavy enough during inflation, there is no isocurvature perturbation.

\begin{figure}[t]
\begin{center}
\includegraphics[scale = 1.0]{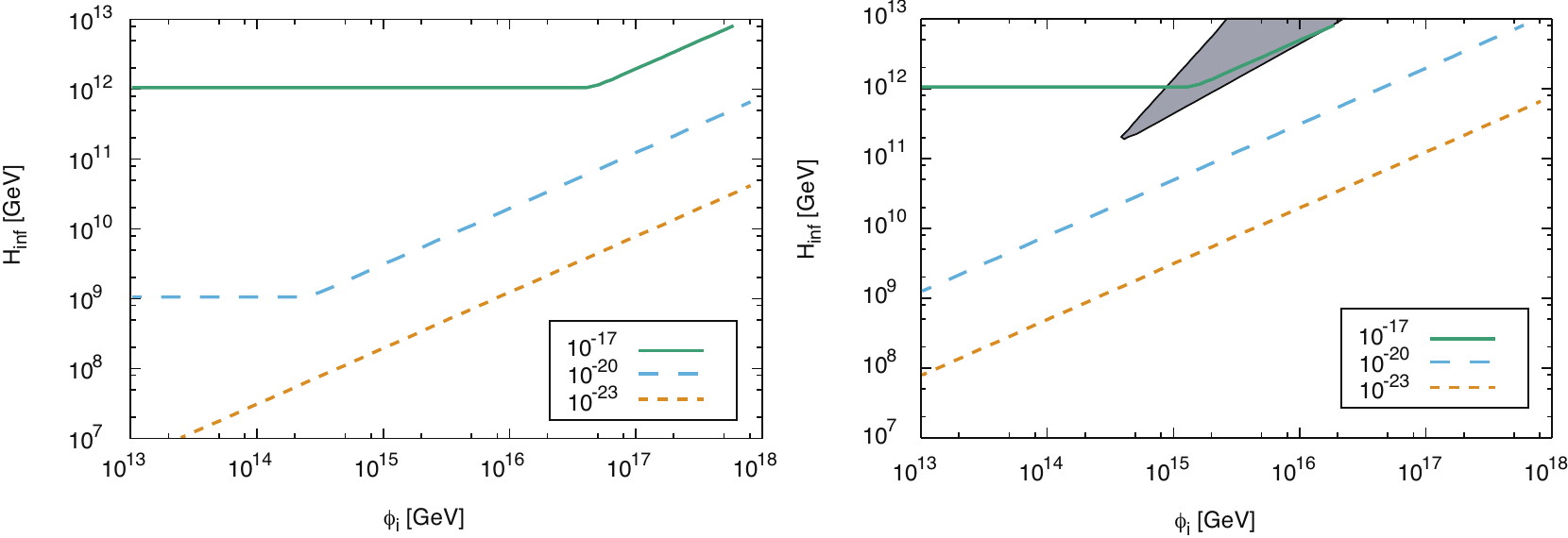}
\end{center}
\caption {\small Contour plot of $Y_\chi$ produced by the inflaton
for $m_\phi=H_{\rm inf}/10$ (left) and $m_\phi=H_{\rm inf}/1000$ (right)
on the plane of $(\phi_i, H_{\rm inf})$.
We fix the decay rate of $\phi$ as $\Gamma_\phi = m_\phi^3 / 128\pi M_P^2$ and
the reheating temperature as $T_\text{R} = 10^{10}\,\text{GeV}$.
The shaded region is excluded due to too large curvature perturbation if $\phi$ remains light during inflation.
}
\label{fig:Ein}
\end{figure}

Fig.~\ref{fig:Ein} shows contours of $Y_\chi \equiv n_\chi/s$ 
produced by the inflaton for $m_\phi=H_{\rm inf}/10$ (left) and $m_\phi=H_{\rm inf}/1000$ 
(right) on the plane of $(\phi_i, H_{\rm inf})$ for $n = 2$ and $w = 1/3$.\footnote{
	One can convert the quantity $\rho_\chi/s = m_\chi Y_\chi$ to the present density parameter 
	$\Omega_\chi = \rho_\chi/\rho_{\rm cr}$ with a critical density $\rho_{\rm cr}$
	through $\Omega_\chi h^2 \simeq 2.8\times 10^8 (m_\chi Y_\chi/{\rm GeV})$
	if $\chi$ is a stable and non-relativistic particle.
	Here $h$ $(\sim 0.7)$ is the present Hubble parameter in units of $100$\,km/s/Mpc.
}
In this plot we have assumed that $\phi$ decays via Planck-suppressed interaction: $\Gamma_\phi \simeq m_\phi^3/(128\pi M_P^2)$.
We have also fixed the reheating temperature after inflation as $T_\text{R} = 10^{10}\,\text{GeV}$.
We can deduce the cosmological constraint mentioned above by multiplying $m_\chi$ as $\rho_\chi/s = m_\chi Y_\chi$ 
for arbitrary value of $m_\chi$ below the inflaton mass.
The shaded region is excluded due to too large curvature perturbation 
if $\phi$ remains light during inflation.
Above mentioned cosmological constraints crucially depend on the mass and lifetime of $\chi$,
which is not fixed in the figure, but we can easily infer $\rho_\chi/s$ (or $\Omega_\chi h^2$)
from these plots once the mass is fixed and compare with various constraints.
One can see that cosmological constrains from the gravitational particle production is rather weak,
so that almost all parameter space is allowed.

%%%%%%%%%%%%%%%%
%%%%%%%%%%%%%%%%

%%%%%%%%%%%%%%%%%%%%%%%%%%%%%%%%%%%%%%%%%%%%%%%%%%
\section{$f(\phi) R$ model}  \label{sec:fphi}
\setcounter{equation}{0}
%%%%%%%%%%%%%%%%%%%%%%%%%%%%%%%%%%%%%%%%%%%%%%%%%%
In this section, we consider gravitational particle production in $f(\phi)R$ models.
One famous example of this class of models is 
the Higgs inflation~\cite{Spokoiny:1984bd,Futamase:1987ua,CervantesCota:1995tz,Bezrukov:2007ep}, 
with $f (\phi) = \xi \phi^2 / M_P^2$.
Interestingly, this coupling, $\xi \phi^2 R / 2$, is inevitably generated 
by radiative corrections~\cite{Herranen:2014cua}.
Here we analyze gravitational particle production for general $f(\phi)R$ models in the Jordan frame.

%%%%%%%%%%%%%%%%%%%%%%%%%%%%%%%%%%%%%%%%%%%%%%%%%%
\subsection{Background dynamics}
%%%%%%%%%%%%%%%%%%%%%%%%%%%%%%%%%%%%%%%%%%%%%%%%%%

Let us consider the action
\begin{align}
	S = \int d^4 x \sqrt{-g}\left( \frac{1}{2}M_P^2 f(\phi)R - \frac{1}{2}(\partial \phi)^2 - V(\phi) + \mathcal L_M \right),
\end{align}
where $\mathcal L_M$ denotes the Lagrangian for matter other than the scalar $\phi$.
The equation of motion of $\phi$ is given by
\begin{equation}
	\ddot \phi + 3H\dot\phi + V' -3M_P^2(2H^2 + \dot H) f'= 0.
\end{equation}
It can be expressed as
\begin{equation}
	\dot \rho_\phi + 3H \left(\rho_\phi + p_\phi\right) - 3M_P^2(2H^2+\dot H)\dot f= 0,  \label{drhophi}
\end{equation}
where $\rho_\phi$ and $p_\phi$ are the same as before.
The Einstein equation reads
\begin{align}
	3H^2f + 3H \dot f &= \frac{\rho_\phi + \rho_M}{M_P^2},  \label{Fried}\\
	 \ddot f + 2H\dot f + (3H^2+2\dot H)f &= -\frac{p_\phi + p_M}{M_P^2}.
\end{align}
By using these equations, we also obtain
\begin{equation}
	\dot \rho_M + 3H(\rho_M + p_M)=0. \label{drhoM}
\end{equation}
Hereafter we again assume that the matter part satisfies the equation of state $p_M = w\rho_M$.
This shows that $\rho_M$ exactly scales as $\rho_M \propto a^{-3(1+w)}$.
The cosmological setup we are considering is the same as the one in the previous section.
We will estimate the oscillating part of the Hubble parameter or the scale factor
induced by the coherent oscillation of $\phi$ which may or may not dominate the universe.
The equation of state of the matter part is taken as a free parameter.

In the following we solve these equations of motion by the following perturbative expansion.
We expand $f(\phi)$ as follows:
\begin{equation}
	f(\phi) \equiv 1 + f_1(\phi) = 1+ c_1\frac{\phi}{M_P} +c_2\frac{\phi^2}{2M_P^2} + \cdots ,  \label{f_exp}
\end{equation}
and regard $f_1$ as a small perturbation.\footnote{
	Again $\phi$ should be regarded as a deviation from the potential minimum $\phi=\phi_{\rm min}$.
	If $\phi_{\rm min}\neq 0$, the first term of (\ref{f_exp}) should be modified as $1-c_1\phi_{\rm min}/M_P$.
}
To be more precise, we require $|\ddot f | \ll H^2$.
Other quantities are also expanded as
\begin{align}
	H &= H_0 + H_1,\\
	a &= a_0 + a_1,\\
	\rho_\phi &= \rho_{\phi0} + \rho_{\phi1},\\
	\rho_M &= \rho_{M0} + \rho_{M1},
\end{align}
where the subscript 0 denotes solutions in $f_1\to 0$ limit, \textit{i.e.}, solutions in the Einstein gravity.
Since $f_1$ directly depends on $\phi$ and hence is a rapidly oscillating function,
quantities such as $H_1, \rho_{\phi1}$, \dots, are also expected to be rapidly oscillating.

Our goal is to express the oscillating part $H_1$, $a_1$,\dots, in terms of $\phi$.
We retain only first order in the oscillating parts 
induced by the non-minimal coupling in the following.
In the equations of motion, the oscillating parts satisfy
\begin{align}
	&2H_0 H_1=- H_0 \dot f_1 - H_0^2f_1+  \frac{\rho_{\phi1} + \rho_{M1}}{3M_P^2},  \label{Fried1}\\
	&\dot \rho_{\phi1} + 3H_1(\rho_{\phi0} + p_{\phi0}) 
	+ 3H_0(\rho_{\phi1} + p_{\phi1}) = 3M_P^2(2H_0^2+\dot H_0)\dot f_1. \label{drhophi1}
\end{align}
Noting that $\dot\rho_{\phi1} \sim \mathcal O(m_\phi \rho_{\phi1})$, 
we can neglect terms of $\sim \mathcal O(H \rho_{\phi1})$ in Eq.~\eqref{drhophi1}.
Then we have
\begin{align}
	\dot \rho_{\phi1} \simeq \left[ 
	3M_P^2(2H_0^2+\dot H_0) + \frac{3}{2}(\rho_{\phi0}+p_{\phi0}) \right] \dot f_1.
\end{align}
This implies $\rho_{\phi1} \sim \mathcal O(\rho_{\rm tot} f_1)$.
Also, $\rho_{M1}$ is suppressed by $m_\phi$ since $\rho_M \propto a^{-3(1+w)}$ is exact.
Thus, by noting that $\dot f_1 \sim \mathcal O(m_\phi f_1)$, 
we find that the second term and the third term in the RHS
of Eq.~\eqref{Fried1} are safely neglected.
As a result, we obtain a simple relation
\begin{equation}
	H_1 \simeq -\frac{\dot f_1}{2} \simeq -\frac{1}{2}\left( c_1\frac{\dot\phi}{M_P}+\cdots \right).  \label{H1}\end{equation}
This is the oscillating part of the Hubble parameter induced by the non-minimal coupling.
Therefore, we arrive at
\begin{align}
	\frac{a_1}{a_0} \simeq 1-\frac{1}{2}f_1,  \label{a1}
\end{align}
and hence we finally find a relation between the oscillating part
of the scale factor and $\phi$.

Eq.~(\ref{H1}) is the same as 
the result obtained from the adiabatic invariant proposed in Ref.~\cite{Ema:2015eqa},
though the proof given in Ref.~\cite{Ema:2015eqa} is applicable 
only to the cases where matter is subdominant.
The point is that there is a so-called ``adiabatic invariant'' $J$:\footnote{
Here integration by parts should be done to remove $\dot{H}$ in the Lagrangian.
}
\begin{align}
	J \equiv -\frac{1}{6M_P^2}\frac{\partial\mathcal L}{\partial H} = \frac{1}{2}(2Hf + \dot f) = \frac{1}{2a^2}\frac{\partial (a^2 f)}{\partial t}.  \label{J}
\end{align}
Here we call a quantity $Q$ as an adiabatic invariant if it satisfies $\dot{Q} = \mathcal{O}(HQ)$.
In the deeply oscillating regime, such a quantity is almost constant
within one oscillation, and hence we can approximately view it as a conserved quantity.
In the Einstein gravity, the Hubble parameter, or equivalently 
the energy density of the scalar field is obviously an adiabatic invariant,
but in extended gravity models it is non-trivial what is the conserved quantity.
Since $J$ is almost constant within one oscillation, 
we can easily extract the oscillation part $H_1$ as
\begin{align}
	H_1 \simeq -\frac{\dot f_1}{2}.
\end{align}
%%

%%%%%%%%%%%%%%%%%%%%%%%%%%%%%%%%%%%%%%%%%%%%%%%%%%
\subsection{Particle production rate}
%%%%%%%%%%%%%%%%%%%%%%%%%%%%%%%%%%%%%%%%%%%%%%%%%%

In the previous subsection, we obtain
\begin{align}
	a(t) \simeq  \langle a (t) \rangle \left( 1 -\frac{c_1}{2}\frac{\phi}{M_P} \right).
\end{align}
This expression is valid up to the first order in $\phi$.
In contrast to the Einstein gravity, here is a linear term in $\phi$ in the
oscillating part of the scale factor, and it induces ``gravitational decay'' of $\phi$.
There also exists quadratic terms 
of the order of $c_1^2$ and $c_2$ 
in addition to the Einstein gravity contribution,
which induce the ``gravitational annihilation'' of $\phi$~\cite{Watanabe:2007tf,Ema:2015dka},
although omitted in this expression.
Note that the gravitational decay is possible only when the non-minimal coupling exists, while the gravitational annihilation
takes place even in the pure Einstein gravity.
Below we consider the production of scalar particles and the graviton.
The production of fermions and gauge bosons is suppressed by their masses and couplings
as we explained before. 

%%%%%%%%%%%%%%%%%%%%%%%%%%%%%%%%%%%%%%%%%%%%%%%%%%
\subsubsection{Scalar}
%%%%%%%%%%%%%%%%%%%%%%%%%%%%%%%%%%%%%%%%%%%%%%%%%%
First let us consider the particle production rate of a scalar coupled with gravity minimally, 
whose action is given by Eq.~\eqref{S_chi}.
The number density of $\chi$ particles produced during one Hubble time 
after $\phi$ begins to oscillate is given by
\begin{equation}
	n_\chi(t) \simeq \frac{C}{32\pi H}\left(\frac{c_1m_\phi^2\Phi}{2M_P}\right)^2.
\end{equation}
It can be interpreted as the decay of $\phi$ into $\chi$ pair with the decay rate
\begin{align}
	\Gamma_{\phi\to\chi\chi} = C\frac{c_1^2}{128\pi} \frac{m_\phi^3}{M_P^2}.  \label{decay_phi}
\end{align}
This decay rate coincides with that calculated in the Einstein frame~\cite{Watanabe:2006ku}.
Contrary to the annihilation case, this effect becomes significant at late time for reasonable choices of $w$ and $n$. 
Noting that each $\chi$ particle has the energy of $m_\phi/2$ at the production, we find 
\begin{align}
	\frac{\rho_\chi (t)}{\rho_\phi(t)} \simeq \frac{C c_1^2 m_\phi^3}{128 \pi M_P^2 H} = \frac{\Gamma_{\phi\to\chi\chi}}{H}.
\end{align}
Thus $\phi$ completely decays into $\chi$ at $H\sim \Gamma_{\phi\to\chi\chi}$
if there is no other decay mode of $\phi$.

%%%%%%%%%%%%%%%%%%%%%%%%%%%%%%%%%%%%%%%%%%%%%%%%%%
\subsubsection{Graviton}  \label{sec:fphi_grav}
%%%%%%%%%%%%%%%%%%%%%%%%%%%%%%%%%%%%%%%%%%%%%%%%%%

Next we apply our formalism to the graviton production.
The graviton action is given by
\begin{align}
	S = \int d\tau d^3x\, a^2(t) f(\phi) \frac{M_P^2}{8}
	\left[ \left(\frac{\partial h_{ij}}{\partial \tau}\right)^2 - (\partial_k h_{ij})^2 \right].
\end{align}
It should be noticed that the $c_1$-dependence vanishes in the overall coefficient $a^2 f(\phi)$.
Hence there is no gravitational decay of $\phi$ into the graviton pair, as opposed to the case of scalar particles~\cite{Ema:2015dka}.
Still there exists a gravitational annihilation of $\phi$ into the graviton pair, which exists even in the Einstein gravity.
The abundance of graviton is similar to Eq.~\eqref{nh_E} 
except for the modification of $\mathcal O(c_2, c_1^2)$.

%%%%%%%%%%%%%%%%%%%%%%%%%%%%%%%%%%%%%%%%%%%%%%%%%%
\subsection{Cosmological implications}  \label{sec:fphi_cos}
%%%%%%%%%%%%%%%%%%%%%%%%%%%%%%%%%%%%%%%%%%%%%%%%%%

In the present case, the contribution from $\phi$ often becomes the dominant one.
The abundance of a massive $\chi$
produced by the gravitational decay of $\phi$ is given by
\begin{align}
	\frac{\rho_\chi}{s} \simeq \Delta' \frac{3m_\chi T_\phi}{2m_\phi}{\rm Br_{\phi\to\chi\chi}} ,
\end{align}
where $T_\phi \sim \sqrt{\Gamma_\phi M_P}$ is the decay temperature of $\phi$
with $\Gamma_\phi$ being the total decay width of $\phi$,
${\rm Br_{\phi\to\chi\chi}} \equiv \Gamma_{\phi\to\chi\chi}/\Gamma_\phi$ is the branching ratio of $\phi$ into $\chi\chi$,
and
\begin{align}
	\Delta' = {\rm min}\left[1,~ \sqrt{H_{\rm dom}/\Gamma_\phi}\right],
\end{align}
which roughly corresponds to the ratio $\rho_\phi/(\rho_\phi + \rho_M)$ at $H = \Gamma_\phi$.
If the gravitational decay is the only decay mode, the branching ratio is $\mathcal O(1)$.
In that case, using Eq.~\eqref{decay_phi}, we obtain
\begin{align}
	\frac{\rho_\chi}{s} \simeq 3\times 10^{-8}\,{\rm GeV}\, \frac{\Delta' c_1}{\sqrt{N+1}}
	 \left( \frac{m_{\phi}}{10^{6}\,{\rm GeV}} \right)^{1/2}
	 \left( \frac{m_{\chi}}{1\,{\rm GeV}} \right),
\end{align}
where we have assumed that there are $N$ light scalar fields other than $\chi$ that thermalize with SM degrees of freedom
and hence ${\rm Br_{\phi\to\chi\chi}} = 1/(N+1)$.
Within the framework of SM, we have $N=4$ corresponding to the four real degrees of freedom of Higgs boson.
Strong constraints are imposed as discussed in Sec.~\ref{sec:cos_grav_E}:
$\rho_\chi/s \lesssim 4\times 10^{-10}\,{\rm GeV}$ if $\chi$ is a non-interacting stable particle, and
$\rho_\chi/s \lesssim 10^{-14}\,{\rm GeV}$ if $\chi$ is a late-decaying particle like moduli.\footnote{
	In the present model, it is likely that $\phi$ obtains a mass of Hubble scale during inflation from $f(\phi) R$ coupling.
	Then there is no DM/dark radiation isocurvature mode even if we consider the DM contribution from inflaton decay.
	Also, $\phi$ is displaced from the minimum of its potential $V (\phi)$ during inflation owing to the $f(\phi) R$ coupling.
	Thus, typically, the initial amplitude is close to the Planck scale,
	unless the potential becomes steeper than the quadratic for a large field value, like $ V \sim \phi^{2n}$ $(n\geq 2)$.
}

Then, suppose that $\chi$ is a (nearly) massless particle such as axion-like particles.
In this case, there is a danger of overproduction of dark radiation.
It is convenient to express the abundance of dark radiation in terms of the effective number of neutrino species
\begin{align}
	\Delta N_\text{eff} = \frac{43}{7}\left( \frac{10.75}{g_{*s}(T_\phi)} \right)^{1/3}\Delta' {\rm Br_{\phi \to \chi\chi}}
	\sim \frac{3\Delta'}{N+1}.
\end{align}
Therefore, if $\phi$ is a dominant component of the universe 
at the decay (i.e. $\Delta'=1$), we may need $N\gtrsim 5$ to satisfy the current constraint
on the dark radiation~\cite{Ade:2015xua}, which is marginal for the SM.\footnote{
Note also that there may be a preference for $\Delta N_{\rm eff} \simeq 0.5$ according to 
a recent observation of the Hubble constant~\cite{Riess:2016jrr}.
}
The bound can be relaxed if $\phi$ has decay modes other than the gravitational decay mode.

\begin{figure}[t]
\begin{center}
\includegraphics[scale = 1.0]{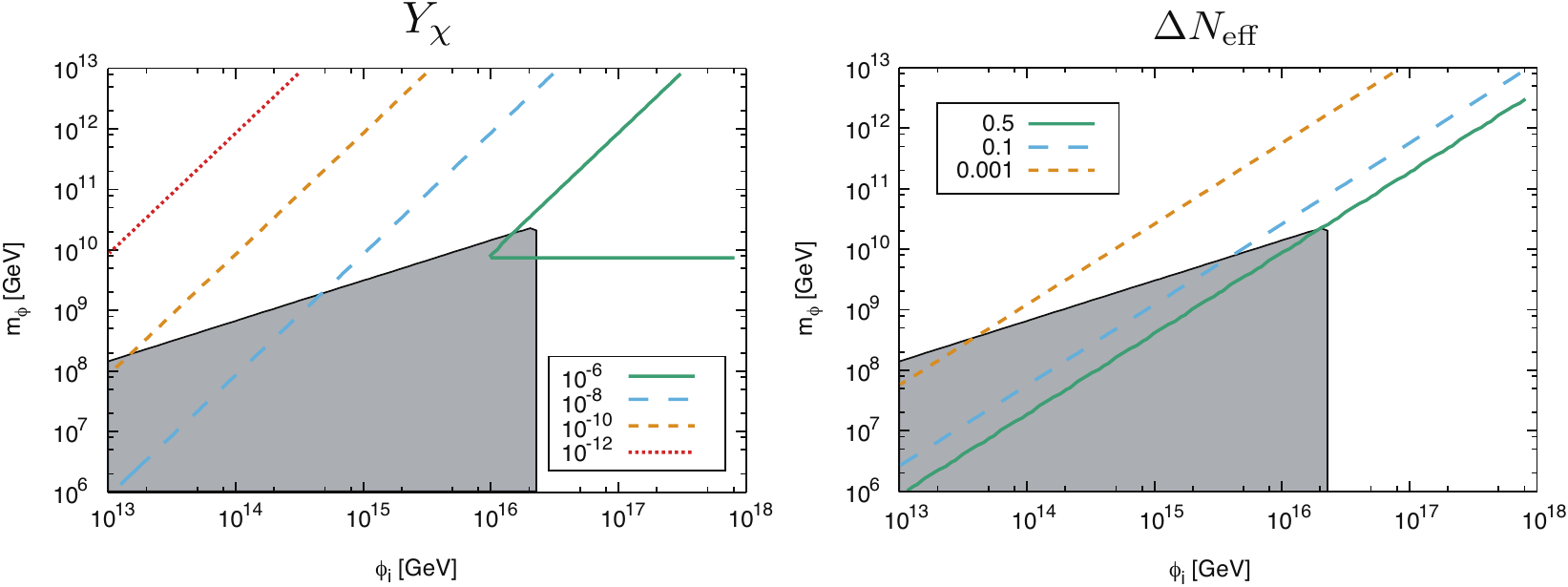}
\end{center}
\caption {\small Contour plot of $Y_\chi$ (left) and $\Delta N_{\rm eff}$ (right)
on the plane of $(\phi_i, m_\phi)$.
The shaded region is excluded due to too large curvature perturbation for $H_{\rm inf}=10^{13}\,$GeV
if $\phi$ remains light during inflation.
}
\label{fig:fphi}
\end{figure}

Fig.~\ref{fig:fphi} shows contours of $Y_\chi$ (left) and $\Delta N_{\rm eff}$ (right) 
produced by $\phi$ for $c_1=1$
on the plane of $(\phi_i, m_\phi)$ for $n = 2$ and $w = 1/3$.
In this plot we have assumed that $\phi$ decays only via the gravitational decay mode
and $N = 4$.
We have also fixed the reheating temperature as $T_\text{R} = 10^{10}\,\text{GeV}$.
The shaded region is excluded due to too large curvature perturbation for $H_{\rm inf}=10^{13}\,$GeV
if $\phi$ remains light during inflation.
Again we emphasize that the cosmological constraints crucially depend on the mass and lifetime of $\chi$.
Comparing with typical constraint (\ref{constraint}), one finds that the a large parameter space of the present scenario is excluded if $\chi$ has long lifetime. 
Of course, the constraints become weaker for small initial amplitude $\phi_i$.

Results presented here are also applied to the inflaton decay by simply regarding $\phi$ as the inflaton and taking $\Delta'=1$.

%%%%%%%%%%%%%%%%
%%%%%%%%%%%%%%%%

%%%%%%%%%%%%%%%%%%%%%%%%%%%%%%%%%%%%%%%%%%%%
\section{$f(R)$ model}
\label{sec:fR}
\setcounter{equation}{0}
%%%%%%%%%%%%%%%%%%%%%%%%%%%%%%%%%%%%%%%%%%%%

In this section, we consider gravitational particle production in $f(R)$ models~\cite{DeFelice:2010aj}.
In $f(R)$ models, there is one additional degree of freedom in the metric sector,
and it induces rapid oscillation of the scale factor.
A famous example is the Starobinsky inflation~\cite{Starobinsky:1980te}, in which a scalar degree of freedom 
causes inflation and reheating~\cite{Vilenkin:1985md,Mijic:1986iv,Gorbunov:2010bn,Arbuzova:2011fu,
Gorbunov:2012ij,Rudenok:2014daa,Terada:2014uia}.
Here we analyze gravitational particle production for general $f(R)$ models in the Jordan frame.

%%%%%%%%%%%%%%%%%%%%%%%%%%%%%%%%%%%%%%%%%%%%
\subsection{Background dynamics}
%%%%%%%%%%%%%%%%%%%%%%%%%%%%%%%%%%%%%%%%%%%%

The action is given by
\begin{align}
	S = \int d^{4}x \sqrt{-g}\,\left(\frac{1}{2}M_{P}^2f(R)+\mathcal L_M\right),
	\label{eq:action_fr}
\end{align}
where $\mathcal L_M$ denotes the Lagrangian for matter.
This model includes one scalar degree of freedom (``scalaron") 
if $F \equiv \partial f/\partial R \neq \rm{const.}$.
The background equations of motion are given by
\begin{align}
	3FH^2 &= \frac{1}{2}\left(FR-f\right) - 3H\dot{F} + \frac{\rho_M}{M_P^2}, 
	\label{eq:eom_fr1} \\
	\ddot{F} - H\dot{F} + 2F\dot{H} &= -\frac{\rho_M + p_M}{M_P^2}.
	\label{eq:eom_fr2}
\end{align}
Note that the second equation
is derived from the first equation just by taking a time derivative
if there is no matter sector.\footnote{
	Recall that the Ricci scalar is given by $R = 6 (\dot H + 2 H^2)$.
}
This is natural because there is only one dynamical degree of freedom, 
\textit{i.e.}, the Hubble parameter, in the gravity sector.
These two equations are combined to yield
\begin{align}
	\ddot F + 3H \dot F + \frac{1}{3}(2f-FR) &= \frac{\rho_M- 3p_M}{3M_P^2},  \label{eqofm_F} \\
	\dot \rho_M + 3H(\rho_M + p_M) &= 0.
\end{align}
Hereafter we assume that the matter satisfies the equation of state $p_M = w \rho_M$, which implies $\rho_M \propto a^{-3(1+w)}$.

In the following, we consider the case where $f(R)$ is given as
\begin{align}
	f(R) = R\left(1 + \frac{c}{n}\left(\frac{R}{M_{P}^{2}}\right)^{n-1}\right)
	\equiv R\left(1 + \frac{cF_1}{n}\right),
\end{align}
where $c$ is a positive constant and $n$\,$( \geq 2)$ is an even integer.
The equation of motion of $F_1$ reads
\begin{equation}
	\ddot {F_1} + 3H \dot {F_1} +\frac{\partial V_{F_1}}{\partial F_1}=0,   
	\label{eqofm_F1}
\end{equation}
where\footnote{
	The potential $V_{F_1}$ is unbounded from below for $n > 2$.
	We only consider the region $c |F_1| \ll 1$ below 
	so that the whole dynamics is described in metastable region.
	Although there can be a quantum tunneling from the metastable vacuum to the deeper minimum,
	we do not discuss it here because higher order terms in $f(R)$ 
	can easily change the structure of the potential for $c|F_1|\gg 1$.
}
\begin{align}
	V_{F_1} =  \frac{n-1}{n}\frac{M_P^2}{3c}|F_1|^{n/(n-1)}\left( 1- \frac{n-2}{2n-1}cF_1 \right) -\frac{1-3w}{3cM_P^2}\rho_M F_1
			+ V_0.   \label{VF1}
\end{align}
Here we have included an $F_1$-independent term $V_0$ to make $V_{F_1}=0$ at the minimum of the potential $F_1=\langle F_1\rangle$:
\begin{align}
	V_0 = \frac{1-3w}{3n c M_P^2}\rho_{M}\langle F_1\rangle.
\end{align}
From Eq.~\eqref{VF1}, we easily find that the minimum of the potential is given by 
\begin{align}
	\langle F_1\rangle = \left(\frac{R_0}{M_P^2}\right)^{n-1},
\end{align}
for $c|F_1|\ll 1$, as expected.
Eq.~\eqref{eqofm_F1} shows that $F_1$ exhibits a similar motion 
to the scalar field under the potential $V_{F_1}$.

Now let us consider the case 
$m_F^2 \equiv \left|(\partial V_{F_1}/\partial F_1)/F_1\right| \gg H^2$, \textit{i.e.}, $F_1$ 
is oscillating rapidly in the effective potential $V_{F_1}$.
We also assume that the inequality
\begin{align}
	c\left\lvert F_1\right\rvert \ll 1,
\end{align}
is satisfied. 
In this case, we can expand quantities as
\begin{align}
	R &= R_0 + R_1,\\
	H &= H_0 + H_{1},\\
	\rho_M &= \rho_{M0} + \rho_{M1},
\end{align}
to the first order in $c$. 
Here quantities with a subscript $0$ corresponds to those in the limit $c\to 0$: 
\begin{align}
	H_0^2 = \frac{\rho_{M0}}{3M_P^2},~~~R_0 = (1-3w)\frac{\rho_{M0}}{M_P^2}.
\end{align}
If there is no matter $(\rho_M=0)$, $F_1$ oscillates around $F_1=0$.
Otherwise, it oscillates around a finite expectation value.
Thus we further divide $F_1$ and $H_1$ into the oscillating part and non-oscillating part as
\begin{align}
	F_1 &= \langle F_1\rangle + \delta F_{1}, \\
	H_1 &= \langle H_1\rangle + \delta H_{1}.  \label{F1}
\end{align}
Note that, for $w=1/3$, $R_0=0$ and hence $ \langle F_1\rangle = 0$.

Our goal is to express the oscillating part of the Hubble parameter or the scale factor
in terms of $F_1$. 
When matter is subdominant, 
the adiabatic invariant $J$ is useful for this purpose.
In App.~\ref{sec:inv}, we show that $J$ is given by
\begin{align}
	J = HF + \frac{\dot{F}}{2},
\end{align}
for the $f(R)$ models.
Thus, by expanding with respect to $c$, we obtain the oscillating part of the Hubble parameter as
\begin{align}
	\delta H_1 \simeq -\frac{c}{2}\dot{\delta F_1}.
	\label{eq:fr_delH1J}
\end{align}
Below we show that this is correct even when matter is non-negligible
by solving the equations of motion directly.
For completeness, we consider both of the cases 
where the scalaron is dominant and subdominant.

For later convenience, 
here we express $H_1$ in terms of $F_1$ by using Eq.~\eqref{eq:eom_fr1}:
\begin{align}
	H_1 \simeq -\frac{1}{2}c\dot{F}_1 
	+ \sqrt{\frac{\rho_{F_1}}{3M_P^2} + \frac{\rho_M}{3M_P^2}
	\left(1- cF_1 + \frac{1}{2}\left(1-3w\right)c\left(F_1 - \frac{1}{n}\left\langle F_1\right\rangle\right) \right) }
	- \sqrt{ \frac{\rho_{M0}}{3M_P^2} },
	\label{eq:fr_gorigori}
\end{align}
where we have kept only leading terms in $c\lvert F_1\rvert$
and defined the ``energy density'' of 
the scalaron $F_1$ as
\begin{align}
	\rho_{F_1} \equiv \frac{3}{2}c^2M_P^2\left( \frac{1}{2}{\dot {F_1}}^2 + V_{F_1} \right).
\end{align}
This is also a non-oscillating quantity.\footnote{
	The contribution to $\rho_{F_1}$ from the constant term $V_0$ is always smaller than $\rho_M$ for $c\langle F_1\rangle \ll 1$.
}
We call this as ``energy density'' because the Hubble parameter 
is given by $H^2 \sim \rho_{F_1}/(3M_P^2)$ for $\rho_{F_1} \gg \rho_M$
as we will show below. Actually, 
$\sqrt{3/2}\,cM_P F_1$ coincides with the canonical scalaron field in the Einstein frame for $c|F_1| \ll 1$.
As one may see from Eqs.~(\ref{eq:fr_delH1J}) and (\ref{eq:fr_gorigori}), 
and as we will see in the following, 
the dominant contribution to the oscillation mode of the Hubble parameter comes from 
the first term in Eq.~(\ref{eq:fr_gorigori}).

%%%%%%%%%%%%%%%%%%%%%%%%%%%%%%%%%%%%%%%%%%%%%%%%%%
\subsubsection{Matter dominated case}
%%%%%%%%%%%%%%%%%%%%%%%%%%%%%%%%%%%%%%%%%%%%%%%%%%

First let us consider the matter\footnote{
	Again, it should not be confused with fluids with the non-relativistic equation of state $w=0$.
	We do not specify $w$ in the following discussion.
} dominated case 
in which $\rho_M \gg \rho_{F_1}$, 
\textit{i.e.}, $c|F_1| \ll 1$ and $c|\dot{\delta F_1}| \ll H_0$. 
For $\rho_{F1} \ll c\lvert F_1\rvert\rho_{M0}$, 
or $\tilde{\delta F_1} \ll \langle F_1\rangle$ with $\tilde{\delta F_1}$ 
being an oscillation amplitude of $\delta F_1$, we obtain
\begin{align}
	\langle H_1\rangle &\sim c \langle F_1\rangle H_0,
\end{align}
and
for $\rho_{F_1} \gg c\lvert F_1\rvert\rho_{M0}$, 
or $\tilde{\delta F_1} \gg \langle F_1\rangle$,
we obtain 
\begin{align}
	\langle H_1\rangle &\sim H_0 \frac{\rho_{F_1}}{\rho_{M0}}.
\end{align}
Note that the latter always holds if $w=1/3$.
In both cases, 
we have
\begin{align}
	\delta H_1 &\simeq -\frac{c}{2}\dot {\delta F_1},
	\label{eq:delH1}
\end{align}
since the oscillating part of $\rho_{M1}$ is 
suppressed by $m_\phi$ as we discussed before.
This implies $|\delta H_1| \ll H_0$ in the matter dominated case.
Thus the scale factor $a$ has also an oscillating part in this model as
\begin{align}
	a (t) \simeq \langle a (t) \rangle \left(1-\frac{c}{2}\delta F_{1}\right).
\end{align}
We can estimate the gravitational particle production rate from this expression.

Let us see the evolution of $\rho_{F_1}$ and $\rho_M$.
From the equation of motion (\ref{eqofm_F1}), we find that $\rho_{F_1}$ scales as $\rho_{F_1} \propto a^{-6n/(3n-2)}$.
It means that the amplitude scales as $\tilde{\delta F_1} \propto a^{-6(n-1)/(3n-2)}$ while $\langle F_1\rangle \propto a^{-3(1+w)(n-1)}$.
Therefore, as time goes on, the relative amplitude of $\tilde{\delta F_1}$ to the mean value $\langle F_1\rangle$ becomes larger for $n > 2(2+w)/3(1+w)$,
which is satisfied for $n \geq 2$ and $w > -1/2$.
It also tends to dominate the universe at a later epoch.
For $n=2$, for example, we have $\rho_{F_1} \propto a^{-3}$ and it scales in the same way as the non-relativistic matter.
Thus, the oscillation energy density will dominate the universe
if the equation of state of background matter is $w>0$.
For $n>2$, $\rho_{F_1}$ decreases more slowly than the non-relativistic matter,
and hence the oscillation energy density eventually dominates the universe
even if $w=0$,
unless $\delta F_1$ decays before the domination due to the production of non-conformally coupled particles as discussed later.

%%%%%%%%%%%%%%%%%%%%%%%%%%%%%%%%%%%%%%%%%%%%%%%%%%
\subsubsection{Oscillation dominated case}
%%%%%%%%%%%%%%%%%%%%%%%%%%%%%%%%%%%%%%%%%%%%%%%%%%

Next let us consider the opposite limit $\rho_M \ll \rho_{F_1}$.
As we have seen above, $\rho_{F_1}$ may eventually dominate the universe 
at a later epoch even if we start with the matter-dominated universe.
In this case, the second term in the potential~\eqref{VF1} can be neglected, and hence 
$F_1$ oscillates around zero: $\tilde{\delta F_1} \gg \langle F_1 \rangle$.
From Eq.~\eqref{eq:fr_gorigori}, we obtain
\begin{align}
	\langle H_1\rangle \simeq \sqrt{\frac{\rho_{F_1}}{3M_P^2}},
\end{align}
hence $\langle H_1\rangle \gg H_0$ and 
\begin{align}
	\delta H_1 \simeq -\frac{c}{2}\dot{\delta F_1}.
\end{align}
This is the same expression as that of the previous case.
The scale factor $a$ can be expressed as
\begin{align}
	a(t) \simeq \langle a(t) \rangle \left(1-\frac{c}{2}\delta F_{1}\right).
\end{align}
In this case we have $|\delta H_1| \sim \langle H_1\rangle \sim H$ and hence the Hubble parameter $H$ violently oscillates.\footnote{
	Although the oscillation amplitude of the Hubble parameter $\delta H_1$ and its averaged value $ \langle H_1\rangle$
	are the same order, $H >0$ is always ensured as is easily checked by solving the Friedmann equation (\ref{eq:eom_fr1}).
}
Similarly to the previous case, from the equation of motion Eq.\eqref{eqofm_F1}, 
we find $\rho_{F_1} \propto a^{-6n/(3n-2)}$
and $\tilde{\delta F_1} \propto a^{-6(n-1)/(3n-2)}$ while $\langle F_1\rangle \propto a^{-3(1+w)(n-1)}$,
hence $\tilde{\delta F_1} \gg \langle F_1\rangle$ is always satisfied 
for $n\geq 2$ and $w > -1/2$ until $\delta F_1$ decays due to particle production discussed later.
The Ricci curvature $R$ oscillates rapidly around $R\sim 0$ and its amplitude decreases as $\tilde R_1\propto a^{-6/(3n-2)}$.
Thus the Hubble parameter scales as $\langle H \rangle \simeq \langle H_1 \rangle \simeq (3n-2)/(3nt) \propto a^{-3n/(3n-2)}$.

In any case, the oscillation of $F_1$, or the oscillation of the scale factor leads to production of non-conformally coupled particles,
and hence it decays.
In the next subsection we estimate the particle production rate.

%%%%%%%%%%%%%%%%%%%%%%%%%%%%%%%%%%%%%%%%%%%%%%%%%%
\subsection{Particle production rate}
%%%%%%%%%%%%%%%%%%%%%%%%%%%%%%%%%%%%%%%%%%%%%%%%%%
In the previous subsection, we obtain
\begin{align}
	a (t) \simeq \langle a (t) \rangle \left(1-\frac{c}{2}\delta F_{1}\right).
\end{align}
As in the case of the $f(\phi)R$ models, 
here is also a linear term in the oscillating part of the scale factor.
Thus ``gravitational decay'' of the scalaron occurs, 
and the scalaron can transfer its energy to other particles efficiently through this process.
We also have terms 
which induce the ``gravitational annihilation'',
although omitted in this expression.
Below we consider the production of minimally coupled scalar particles and the graviton.
The production of fermions and gauge bosons is again suppressed by their masses and couplings.

%%%%%%%%%%%%%%%%%%%%%%%%%%%%%%%%%%%%%%%%%%%%%%%%%%
\subsubsection{Scalar}
%%%%%%%%%%%%%%%%%%%%%%%%%%%%%%%%%%%%%%%%%%%%%%%%%%

First let us consider the particle production rate of minimally coupled scalar, 
whose action is given by Eq.~\eqref{S_chi}.
By noting that $\ddot a/a \simeq -(c/2) m_F^2 F_1$ and using Eq.~\eqref{nchi_grav}, 
we obtain the number density of $\chi$ created in one Hubble time as
\begin{align}
	n_\chi(t) \simeq \frac{C \tilde{R}^2}{1152\pi H},
\end{align}
where $\tilde{R}$ is the amplitude of the Ricci scalar $R$.
This expression does not depend on $n$ except for the small dependence in the $\mathcal O(1)$ constant $C$.
From this we can read off the effective ``decay rate'' of $F_1$ as
\begin{align}
	\Gamma_{F_1\to\chi\chi} = \frac{C}{384\pi} \frac{n}{n-1}\frac{m_F^3}{M_P^2},
\end{align}
which coincides with the decay rate of 
a canonical scalaron field calculated in the Einstein frame~\cite{Gorbunov:2010bn}.
The ratio of the energy density of the created particles in each Hubble time 
to the scalaron energy density is given by
\begin{align}
	\frac{\rho_\chi(t)}{\rho_{F_1}(t)} \simeq \frac{C}{384\pi} \frac{n}{n-1}\frac{m_F^3}{M_P^2 H} = \frac{\Gamma_{F_1\to\chi\chi}}{H}.
\end{align}
This ratio becomes $\mathcal O(1)$ at some epoch even if it is initially much smaller
since $m_F$ is an increasing function of time.
At that time, $\delta F_1$ completely ``decays'' into $\chi$ particles.
If $\rho_{F_1}$ dominates the universe, it corresponds to the completion of the reheating.
Actually if $\chi$ is the SM Higgs boson, they are thermalized soon.

%%%%%%%%%%%%%%%%%%%%%%%%%%%%%%%%%%%%%%%%%%%%%%%%%%
\subsubsection{Graviton}
%%%%%%%%%%%%%%%%%%%%%%%%%%%%%%%%%%%%%%%%%%%%%%%%%%

The graviton action is given by
\begin{align}
	S = \int d\tau d^3x\, a^2(t) F(R) \frac{M_P^2}{8}
	\left[ \left(\frac{\partial h_{ij}}{\partial \tau}\right)^2 - (\partial_k h_{ij})^2 \right].
\end{align}
It is the combination $a^2(t) F(R)$ that determines the graviton production rate.
It is estimated as
\begin{align}
	a^2(t) F(R) \simeq a_0^2\left( 1+ \mathcal O\left( c^2 F_1^2 \right) \right).
\end{align}
Note that, similarly to the case of $f(\phi)R$ models, the linear term in $c\delta F_1$ vanishes, hence there is no ``decay'' of $F_1$ into the graviton pair.
Compared with the scalar, the graviton abundance is suppressed by $c^2F_1^2$:
\begin{align}
	n_h(t) \simeq \frac{C (cF_1\tilde{R})^2}{1152\pi H}.
\end{align}
The graviton production becomes less efficient as time goes on
due to the time-dependent suppression factor $F_1^2$.\footnote{
	This is inconsistent with Ref.~\cite{Schiappacasse:2016nei}.
	Probably they did not take into account $F(R)$ appearing in front of the graviton kinetic term.
}
It corresponds to the gravitational annihilation of the oscillating scalaron field in the Einstein frame interpretation as written in Sec.~\ref{sec:grav_E}.

%%%%%%%%%%%%%%%%%%%%%%%%%%%%%%%%%%%%%%%%%%%%%%%%%%
\subsection{Cosmological implications}
%%%%%%%%%%%%%%%%%%%%%%%%%%%%%%%%%%%%%%%%%%%%%%%%%%

Let us discuss cosmological implications of gravitational particle production in $f(R)$ models.
To be concrete, we take $n=2$ in the following.
If there is no matter initially, it can cause successful Starobinsky inflation, 
but here we concentrate on the cases where the inflation occurs in some other sector and
$F_1$ oscillation begins after inflation, which later becomes dominant or subdominant component of the total energy density.

The effects of gravitational particle production in the $f(R)$ model is similar to the case of $f(\phi)R$ model with $c_1\neq 0$
studied in Sec.~\ref{sec:fphi_cos}.
The massive $\chi$ abundance produced by the gravitational $F_1$ decay is given by
\begin{align}
	\frac{\rho_\chi}{s} \simeq \Delta' \frac{3m_\chi T_F}{2m_F}{\rm Br}_{F_1\to\chi\chi} ,
\end{align}
where $T_F \sim \sqrt{\Gamma_{F_1} M_P}$ is the decay temperature of $F_1$
and ${\rm Br}_{F_1\to\chi\chi} \equiv \Gamma_{F_1\to\chi\chi}/\Gamma_{F_1}$ is the branching ratio of $F_1$ into $\chi\chi$
with $\Gamma_{F_1}$ being the total decay width of $F_1$ and
\begin{align}
	\Delta' = {\rm min}\left[1,~ \sqrt{H_{\rm dom}/\Gamma_{F_1}}\right].
\end{align}
Here $H_{\rm dom}$ is the Hubble parameter at which $F_1$ would dominate the universe
and $\Delta'$ roughly corresponds to the ratio $\rho_{F1}/(\rho_{F_1} + \rho_M)$ at $H = \Gamma_{F_1}$.
Writing the initial condition of $F_1$ as $F_{1i}$, we obtain
$H_{\rm dom} = \Gamma_{\rm inf}(F_{1i}^2/6M_P^2)^2$ for $m_F > \Gamma_{\rm inf}$ and
$H_{\rm dom} = m_F(F_{1i}^2/6M_P^2)^2$ for $m_F < \Gamma_{\rm inf}$, respectively.
The energy density of $\chi$ is then given by
\begin{align}
	\frac{\rho_\chi}{s} \simeq 2\times 10^{-8}\,{\rm GeV}\, \frac{\Delta'}{\sqrt{N+1}}
	 \left( \frac{m_{F}}{10^{6}\,{\rm GeV}} \right)^{1/2}
	 \left( \frac{m_{\chi}}{1\,{\rm GeV}} \right).
\end{align}
This is severely constrained if $\chi$ is a stable non-interacting particle, or it is a late decaying moduli
as shown in Sec.~\ref{sec:cos_grav_E}.
In our setup, $F_1$ remains light during inflation and it obtains long-wavelength quantum fluctuation.
Whether $\chi$ has (large) isocurvature perturbation or not depends on the dominant source of the curvature perturbation:
if it is the inflaton, the fluctuation of $\chi$ is mostly uncorrelated isocurvature and cannot be a dominant DM,
while if it is $F_1$, there is essentially no isocurvature mode except for (small) contribution from the inflaton oscillation. Also there is no isocurvature mode if $F_1$ itself is the inflaton.

If $\chi$ is a practically massless non-interacting particle, we have
\begin{align}
	\Delta N_\text{eff} = \frac{43}{7}\left( \frac{10.75}{g_{*s}(T_F)} \right)^{1/3}\Delta' {\rm Br}_{F_1\to\chi\chi}
	\sim \frac{3\Delta'}{N+1}.
\end{align}
Again we have a stringent constraint.\footnote{
	Dilaton dark radiation from the decay of scalaron field in $R^2$ model was discussed in Ref.~\cite{Gorbunov:2013dqa}
}
Constraints are similar to the case of $f(\phi)R$ model with $c_1=1$ after $\phi_i$ is replaced with $F_{1i}$
and readers are referred to Fig.~\ref{fig:fphi}.

These results can  be applied to the reheating of the Starobinsky inflation model once we take $\Delta'=1$
and $m_F \simeq 3\times 10^{13}\,$GeV.
It is noticeable that in the Starobinsky model with a minimal extension of an axion,
we have $N=4$ (corresponding to the SM Higgs boson) 
and the axion dark radiation may be detectable in future CMB experiment.\footnote{
	If the radial component of the Peccei-Quinn scalar is lighter than the inflaton, we have $N=5$.
	But it dominantly decays into the axion pair, and the axion dark radiation becomes even more abundant.
}
Such axion dark radiation can also have isocurvature mode depending on the origin of dominant curvature perturbation.

%%%%%%%%%%%%%%%%%%%%%%%%%%%%%%%%%%%%%%%%%%%%
\section{$G^{\mu\nu}\partial_\mu\phi \partial_\nu\phi$ model}
\label{sec:G}
\setcounter{equation}{0}
%%%%%%%%%%%%%%%%%%%%%%%%%%%%%%%%%%%%%%%%%%%%

Finally, we study a scalar field with 
a non-minimal derivative coupling to gravity, namely 
$\mathcal L\sim G^{\mu\nu}\partial_\mu\phi \partial_\nu\phi$ with $G^{\mu\nu}$ being the Einstein tensor.
An example with such a coupling is the new Higgs inflation model~\cite{Germani:2010gm}.
This class of model has an advantage that it does not introduce an additional
degree of freedom although the action itself contains higher derivatives.
In fact, it is the simplest version of the $G_5$-type (or the $G_4$-type involving the kinetic term) 
models in the context of 
the Horndeski or generalized Galileon theories~\cite{Horndeski:1974wa,Deffayet:2011gz,Kobayashi:2011nu}.

%%%%%%%%%%%%%%%%%%%%%%%%%%%%%
\subsection{Background dynamics}
%%%%%%%%%%%%%%%%%%%%%%%%%%%%%

We consider the following action
\begin{align}
	S = \int d^4x \sqrt{-g}\left[ \frac{1}{2}M_P^2 R - \frac{1}{2}\left(g^{\mu\nu}-\frac{G^{\mu\nu}}{M^2} \right)\partial_\mu\phi\partial_\nu\phi - V(\phi) + \mathcal L_M \right].
\end{align}
The background equation of motion of $\phi$ is given by
\begin{align}
	\left(1+ \frac{3H^2}{M^2} \right)\ddot\phi + 3H\left(1+\frac{3H^2 + 2\dot H}{M^2} \right)\dot \phi + V' = 0.
\end{align}
The Friedmann equation reads
\begin{align}
	3H^2= \frac{\rho_\phi + \rho_M}{M_P^2},~~~~&\rho_\phi\equiv \left(1+ \frac{9H^2}{M^2} \right)\frac{\dot\phi^2}{2} + V, \label{Fried_G}  \\[.5em]
	3H^2+2\dot H = -\frac{p_\phi+p_M}{M_P^2}, ~~~~&p_\phi\equiv\left(1- \frac{3H^2}{M^2} \right)\frac{\dot\phi^2}{2}-V-\frac{1}{M^2}\frac{d}{dt}\left(H\dot\phi^2\right),
\end{align}
where $\rho_M$ and $p_M$ are the same as before.
From these equations, we obtain
\begin{align}
	&\dot\rho_\phi + 3H(\rho_\phi + p_\phi)=0,  \label{rhophi_G} \\
	&\dot\rho_M + 3H(\rho_M + p_M)=0.  \label{rhoM_G}
\end{align}
The oscillating regime of this system without the matter $(\rho_M=0)$ 
was extensively studied in Refs.~\cite{Jinno:2013fka,Ema:2015oaa}.
It is found that this system has a so-called gradient instability 
in the oscillating epoch if $\phi$ dominates the universe
and the non-minimal kinetic term dominates over the standard one $(H\gtrsim M)$~\cite{Ema:2015oaa}.\footnote{
	Refs.~\cite{Germani:2015plv,Myung:2016twf} argued a subtlety on the gauge choice $\delta\phi=0$ around the end point of the field oscillation $\dot\phi=0$
	in analyzing the perturbation of the scalar field oscillation.
	It does not matter, however, for the discussion here.
	This is because the gradient instability occurs in the time scale much shorter than the one scalar oscillation period:
	the relevant wavenumber for the instability is $|c_s|k \gg m_\text{eff}$ with $c_s$ being the sound speed.
}
The gradient instability indicates that the sound speed squared of the scalar perturbation
becomes negative for a finite period during one oscillation, which means that the scalar fluctuations are exponentially enhanced.
In particular, the enhancement rate is larger for higher momentum modes.
The system soon becomes non-linear, and it is quite difficult to follow the dynamics at least analytically.
In order to avoid this instability when $\phi$ dominates the universe, 
the non-minimal kinetic term must be so small that the model
effectively reduces to just a canonical scalar field with Einstein gravity.
Therefore, we limit ourselves to the case where $\rho_M$ dominates the universe and 
the $\phi$-oscillation is a subdominant component, and also require that there is no gradient instability.

For later convenience, we define the effective mass of the scalar as
\begin{align}
	m_\text{eff} \equiv {\rm min}\left[ 1, \frac{M}{H} \right] \times\left. \sqrt{\frac{V'}{\phi}}\right|_{\phi=\Phi},
\end{align}
where $\Phi$ denotes the oscillation amplitude of $\phi$. This roughly corresponds to the scalar oscillation frequency.
Note that the energy conservation (\ref{rhophi_G}) immediately means that
\begin{align}
	\dot\rho_\phi \sim \begin{cases}
		m_\text{eff} \rho_\phi &{\rm for}~~~H\gtrsim M,\\
		\frac{H^2}{M^2}m_\text{eff} \rho_\phi &{\rm for}~~~ M^2/m_\text{eff} \lesssim H \lesssim  M,\\
		H \rho_\phi &{\rm for}~~~H \lesssim M^2/m_\text{eff}.
	\end{cases}
\end{align}
This implies that $\rho_\phi$ is a rapidly oscillating quantity for $H\gg M$.
The relative amplitude of the oscillating part of $\rho_\phi$ is estimated as
\begin{align}
	\frac{\delta\rho_\phi}{\rho_{\phi0}} \sim \begin{cases}
		\mathcal O(1) &{\rm for}~~~H\gtrsim M,\\
		\mathcal O\left(\frac{H^2}{M^2}\right) &{\rm for}~~~ M^2/m_\text{eff} \lesssim H \lesssim  M,\\
		\mathcal O\left(\frac{H}{m_\text{eff}}\right) &{\rm for}~~~H \lesssim M^2/m_\text{eff}.
	\end{cases}
	\label{eq:delrho_G5}
\end{align}
The last case is the same as that of the canonical scalar with the Einstein gravity.
On the other hand, as usual, $\rho_M$ just scales as $a^{-3(1+w)}$ and therefore its relative oscillation amplitude is small:
$\delta\rho_M / \rho_M \sim \delta H/m_\text{eff}$. 
Therefore the oscillating part of the Hubble parameter is expressed as
\begin{align}
	\frac{\delta H}{H_0} \simeq\frac{\delta\rho_{\phi}}{2\rho_{M}}.
	\label{eq:delH_G5}
\end{align}
The scale factor can also be expanded as
\begin{align}
	a(t) \simeq a_0 \left( 1 + \mathcal O\left( \frac{\delta H}{m_\text{eff}} \right) \right).
\end{align}
We can calculate the particle production using these expressions.

Let us make an order estimation on the condition to avoid the gradient instability.
The sound speed squared of the scalar field is given by~\cite{Ema:2015oaa}\footnote{
	It can be estimated as $c_s^2\sim (1-a^2G^{ij}/M^2)/(1+G^{00}/M^2)$
	with $G^{00}=3H^2$ and $G^{ij} = -a^{-2}(3H^2 + 2\dot H)\delta^{ij}$.
}
\begin{align}
	c_s^2 \sim \begin{cases}
	1 + \mathcal O\left(\frac{\dot H}{H^2}\right)&~~~{\rm for}~~~H \gg M, \\[.5em]
	1 + \mathcal O\left(\frac{\dot H}{M^2}\right)&~~~{\rm for}~~~H \ll M.
	\end{cases}
\end{align}
If $H$ is violently oscillating, the sound speed squared may be negatively large, 
which leads to a gradient instability.
Thus we require ${\rm min}\left[|\dot H/H^2|,\, |\dot H/M^2|\right] \lesssim 1$ to avoid the instability.
From Eq.~(\ref{Fried_G}), one can see that this condition is written as
\begin{align}
	{\rm min}\left[1, \frac{H^4}{M^4}\right]
	\frac{m_\text{eff}}{H} \frac{\rho_\phi}{\rho_M} \lesssim 1~~~{\rm for}~~~M^2/m_\text{eff} \lesssim H.
	\label{eq:cond_inst}
\end{align}
No condition is required for $H \lesssim M^2/m_\text{eff}$.
Hereafter we assume that this inequality is always satisfied.
From this expression it is clear that if $\phi$ is the dominant component of the universe, we must have $H\ll M$
to avoid the instability, as stated above.

Since we have imposed the condition~\eqref{eq:cond_inst}, 
the evolution of $\phi$ is greatly simplified.
By noting $\dot H\simeq -3(1+w)H^2/2$, the equation of motion is approximated as
\begin{align}
	\ddot \phi - 3Hw \dot\phi + \frac{M^2}{3H^2}V'=0~~~{\rm for}~~~H\gg M.
\end{align}
For $H\ll M$, the equation of motion is the same as that of the canonical scalar field.
Using the Virial theorem, we find
\begin{align}
	\Phi \propto \begin{cases}
		a^{-3(1-w)/(n+2)}&{\rm for}~~~H\gg M, \\
		a^{-6/(n+2)} &{\rm for}~~~H\ll M,
	\end{cases}
\end{align}
where we have assumed $V\propto \phi^n$.

%%%%%%%%%%%%%%%%%%%%%%%%%%%%%%%%%%%%%%%%%%%%%%%%%%
\subsection{Particle production rate}
%%%%%%%%%%%%%%%%%%%%%%%%%%%%%%%%%%%%%%%%%%%%%%%%%%
In the previous subsection, we have seen that
\begin{align}
	a(t) \simeq \langle a(t) \rangle \left( 1 + \mathcal{O}\left(\frac{\delta H}{m_\text{eff}}\right) \right),
\end{align}
with the oscillation part of the Hubble parameter $\delta H$ 
given by Eqs.~\eqref{eq:delrho_G5} and~\eqref{eq:delH_G5}.
Thus we can view the particle production
as the gravitational annihilation in the present case.
In addition, there is a direct coupling between the graviton and the scalar field
induced by the non-minimal derivative coupling to gravity, and it can also cause the graviton production.
Below we analyze the gravitational particle production of a minimally coupled scaler field and graviton.
We do not discuss fermions and vector bosons 
since they are classically Weyl-invariant in the massless limit.
%%%%%%%%%%%%%%%%%%%%%%%%%%%%%%%%%%%%%%%%%%%%%%%%%%
\subsubsection{Scalar}
%%%%%%%%%%%%%%%%%%%%%%%%%%%%%%%%%%%%%%%%%%%%%%%%%%

Now we evaluate the production rate of a minimally coupled scalar~\eqref{S_chi}.
The number density of the produced particles per one Hubble time is estimated by 
Eq.~\eqref{nchi_grav} as
\begin{align}
	n_\chi(t) \sim \begin{cases}
		\displaystyle
		\frac{C}{512\pi H} \left(\frac{Hm_\text{eff}^3\Phi^2}{M^2 M_P^2}\right)^2&{\rm for}~~~ M^2/m_\text{eff} \lesssim H,\\[1.em]
		\displaystyle
		\frac{C}{512\pi H}\left(\frac{m_\text{eff}^2\Phi^2}{M_P^2}\right)^2 &{\rm for}~~~H \lesssim M^2/m_\text{eff}.
	\end{cases}
	\label{nchi_G}
\end{align}
From this we can deduce the effective ``annihilation rate'' of $\phi$ into $\chi$ pair, as
\begin{align}
	\Gamma_{\phi\phi\to\chi\chi} \sim \begin{cases}
		\displaystyle
		\frac{C}{512\pi} \frac{\Phi^2 m_\text{eff}^5}{M^2 M_P^4} &{\rm for}~~~H\gtrsim M,\\[1em]
		\displaystyle
		\frac{C}{512\pi} \frac{H^2\Phi^2 m_\text{eff}^5}{M^4 M_P^4} &{\rm for}~~~ M^2/m_\text{eff} \lesssim H \lesssim  M,\\[1em]
		\displaystyle
		\frac{C}{512\pi} \frac{\Phi^2 m_\text{eff}^3}{M_P^4} &{\rm for}~~~H \lesssim M^2/m_\text{eff}.
	\end{cases}
\end{align}
To obtain these results,
we have defined the number density of $\phi$ as $n_\phi \equiv \rho_\phi/m_\text{eff}$.
It is soon realized that $\Gamma_{\phi\phi\to\chi\chi} / H$ is non-decreasing function of time during $H\gtrsim M$ 
for $5/2 + 9/2(1+6w) \geq n$ (\textit{i.e.}, $2 \leq n \leq 4$ for $w \leq 1/3$).
It is easily shown that $\Gamma_{\phi\phi\to\chi\chi}$ never exceeds $H$ under the condition~\eqref{eq:cond_inst}.

%%%%%%%%%%%%%%%%%%%%%%%%%%%%%%%%%%%%%%%%%%%%%%%%%%
\subsubsection{Graviton}
%%%%%%%%%%%%%%%%%%%%%%%%%%%%%%%%%%%%%%%%%%%%%%%%%%

For the graviton production,
in addition to the ``usual'' gravitational production similar to Eq.~\eqref{nchi_G}, 
there is a contribution coming from
the direct coupling between $\phi$ and the graviton through the non-minimal kinetic term.
The former is the same as that of the scalar field, and hence we concentrate
on the latter here.
As shown in Ref.~\cite{Ema:2015oaa}, the graviton kinetic term is written as
\begin{align}
	S \sim \int dt' d^3x\,\frac{1}{2}\sqrt{ 1 - \left(\frac{(d\phi/dt')^2}{2M_P^2M^2}\right)^2}\left[ 
	\left( \frac{\partial h_{ij}}{\partial t'} \right)^2- (\partial_l h_{ij})^2 \right],
\end{align}
where
\begin{align}
	dt' \equiv \left (\frac{1+\frac{\dot\phi^2}{2M^2M_P^2}}{1-\frac{\dot\phi^2}{2M^2M_P^2}}\right)^{1/2}dt,
\end{align}
and we have omitted the scale factor here.
The effective annihilation rate of $\phi$ into the graviton pair is
\begin{align}
	\frac{\Gamma_{\phi\phi\to hh}}{\Gamma_{\phi\phi\to\chi\chi}} \sim \begin{cases}
		\displaystyle
		\left(\frac{m_\text{eff}}{H}\right)^2 
		\left(\frac{\rho_\phi}{\rho_M}\right)^2
		&{\rm for}~~~H\gtrsim M,\\[1em]
		\displaystyle
		\frac{H^4}{M^4}
		\left(\frac{m_\text{eff}}{H}\right)^2 
		\left(\frac{\rho_\phi}{\rho_M}\right)^2
		&{\rm for}~~~M^2/m_\text{eff} \lesssim H \lesssim M,\\[1em]
		\displaystyle
		\left(\frac{H}{M^2/m_\text{eff}}\right)^4
		\left(\frac{\rho_\phi}{\rho_M}\right)^2
		&{\rm for}~~~H \lesssim M^2/m_\text{eff}.
	\end{cases}
\end{align}
Therefore, this annihilation mode cannot exceed the ordinary gravitational production
if we prohibit the gradient instability.\footnote{
	If we allow the gradient instability to occur, the graviton (or gravitational wave) signal would be 
	much more stronger, although the precise analysis is difficult to perform.
}

%%%%%%%%%%%%%%%%%%%%%%%%%%%%%%%%%%%%%%%%%%%%%%%%%%
\subsection{Cosmological implications}
%%%%%%%%%%%%%%%%%%%%%%%%%%%%%%%%%%%%%%%%%%%%%%%%%%

%%%%%%%%%%%%%%%%
\begin{figure}[t]
\begin{center}
\includegraphics[scale = 1.0]{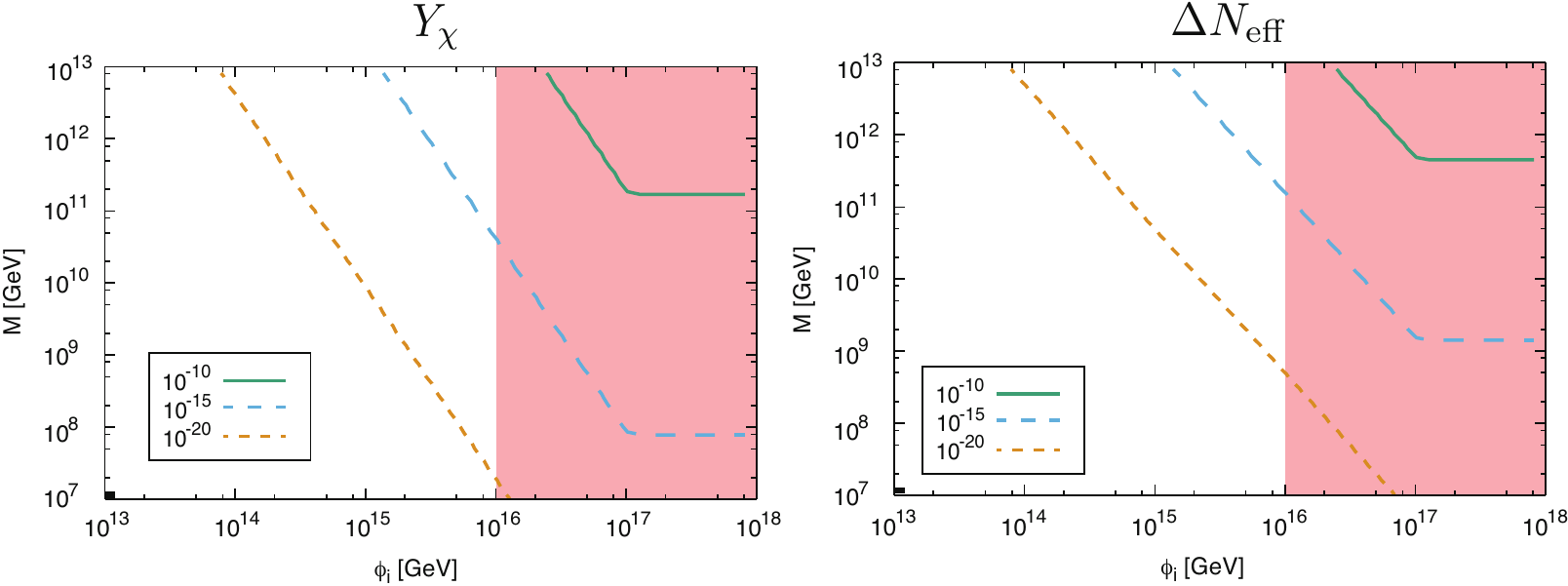}
\end{center}
\caption {\small Contour plot of $Y_\chi$ (left) and $\Delta N_{\rm eff}$ (right) for $m_\phi=100M$
on the plane of $(\phi_i, M)$.
In the shaded region there is a gradient instability.
}
\label{fig:M}
\end{figure}
%%%%%%%%%%%%%%%%

Now we discuss the cosmological implications of the gravitational particle production.
To be concrete, we take $n=2$ and $w=1/3$.
The dominant contribution to the abundance of the minimally-coupled scalar comes from $H\sim M$,
since $\Gamma_{\phi\phi\to \chi\chi}/H$ is an increasing function of time for $H\gtrsim M$ while it is decreasing at $H\lesssim M$.
We obtain
\begin{align}
	\left.Y_\chi\right\rvert_{H > M} 
	\sim \frac{\alpha m_\text{eff}^2 T}{M_P^2 H}\left( \frac{\rho_\phi}{\rho_M} \right)^2,
\end{align}
for $H\gtrsim M$, where $T \sim \rho_M/s$ is the ``temperature''
of the universe and $\alpha\sim 10^{-3}$ is a numerical coefficient.
As an extreme case, let us assume that the inequality (\ref{eq:cond_inst}) is almost saturated at $H\sim M$.
Then we have
\begin{align}
	\frac{\rho_\chi}{s} \lesssim \frac{\alpha m_\chi M^{3/2}}{M_P^{3/2}} 
	\sim 3\times10^{-10}\,{\rm GeV} \left(\frac{\alpha}{10^{-3}}\right)
	\left( \frac{m_\chi}{10^{6}\,{\rm GeV}} \right)
	\left( \frac{M}{10^{10}\,{\rm GeV}} \right)^{3/2}.
\end{align}
The observational upper bound is $\rho_\chi/s \lesssim 4\times 10^{-10}$\,GeV for a stable non-interacting $\chi$ field
and $\rho_\chi/s \lesssim 10^{-14}$\,GeV for $ \chi$ as massive moduli.
Note again that if $\phi$ remains light during inflation, $\chi$ particle produced in this way has isocurvature fluctuation
and hence cannot be a dominant component of DM.
In the present model, $\phi$ cannot dominantly contribute to the curvature perturbation because 
the energy density of $\phi$ must be sufficiently small to avoid the gradient instability, and such a subdominant curvaton would lead to too large non-Gaussianity.

The graviton abundance is also the same as that of the light scalar field.
The corresponding peak frequency is estimated as
\begin{align}
	f_{\rm GW} \sim 2\times 10^9\,{\rm Hz}\,\left( \frac{m_\phi}{10^{13}\,{\rm GeV}} \right)
	\left( \frac{10^{10}\,{\rm GeV}}{M} \right)^{1/2}.
\end{align}
Around this frequency range, the gravitational wave abundance is too small to detect.

Fig.~\ref{fig:M} shows contours of $Y_\chi$ (left) and $\Delta N_{\rm eff}$ (right) for $m_\phi=100M$
on the plane of $(\phi_i, M)$ for $n=2$ and $w=1/3$.
We have implicitly assumed that the inflation scale $H_\text{inf}$ satisfies
$H_\text{inf} > (M/H_\text{inf}) m_\phi$ ($H_\text{inf} > 10 M$ for $m_\phi = 100M$)
and $\phi$ decays into radiation after $H \sim M$,
but before the domination.
In the red shaded region there is a gradient instability.
From this figure, it is seen that once we avoid the gradient instability, which would otherwise invalidate
the reheating analysis, cosmological constraints are not so stringent (compare with typical constraint for massive long-lived particle (\ref{constraint})).

%%%%%%%%%%%%%%%%%%%%%%%%%%%%%%%%%%%%%%%%%%%%
\section{Conclusions and discussion}   \label{sec:con}
%%%%%%%%%%%%%%%%%%%%%%%%%%%%%%%%%%%%%%%%%%%%

In this paper, we have studied the gravitational particle production caused by 
a coherently oscillating scalar field in the universe.
We have treated the Einstein gravity, $f(\phi)R$ gravity, $f(R)$ gravity and 
$G^{\mu\nu}\partial_{\mu}\phi\partial_{\nu}\phi$ gravity theories where $\phi$ is the scalar field,
$R$ is the Ricci scalar and $G^{\mu\nu}$ is the Einstein tensor, respectively.

We have estimated the particle production rate 
for such a broad class of models in a unified framework.
In particular, we pay attention to an oscillating part of the scale factor,
which makes manifest how the background oscillation produces non-conformally coupled particles.
A coherently oscillating scalar field, no matter whether it is dominant or subdominant,
induces an oscillating feature of the scale factor. 
It exists even in the Einstein gravity theory,
and is more violent for the extended gravity theories.
All particles couple to the scale factor unless they are Weyl-invariant, 
and feel the oscillation of the scale factor.
Thus gravitational particle production by the scalar field occurs through its oscillation.
In the previous paper~\cite{Ema:2015dka}, we considered only the case
where the scalar field dominates the universe.
In this paper, we have extended our study 
so that 
it can be applied to a subdominant scalar field as well.
We have also treated a broader class of gravity theories systematically.
For the Einstein gravity theory, the production caused by the inflaton 
is larger than any other subdominant scalar fields.
However, in the extended gravity theories,
the contribution from the subdominant scalar field, other than inflaton, 
can be the dominant one.

An interesting feature of our viewpoint is that, 
once we express the oscillating part of
the scale factor by the coherently oscillating scalar field, 
we can easily deduce effective couplings between the scalar field 
and other particles mediated by the gravity from the Lagrangian.
In the Einstein and $G^{\mu\nu}\partial_{\mu}\phi\partial_{\nu}\phi$ theories,
the oscillating part of the scale factor depends quadratically on 
the scalar field, and hence we can view it as ``gravitational annihilation."
In the $f(\phi)R$ and $f(R)$ theories, 
it depends linearly on the scalar field in general, 
and hence we can view it as ``gravitational decay.''
We can easily estimate the production rate, which coincides with that
obtained from more rigorous calculations.
For example, in our viewpoint,
it is clear that the scalar field (or the scalaron) does not decay
into the gravitons in the $f(\phi)R$ and $f(R)$ theories.
Indeed we have explicitly seen that the direct coupling cancels with the oscillating part
of the scale factor, resulting in no effective coupling between the scalar field and the graviton.
It is consistent with the results in the Einstein frame.

We have also discussed the cosmological implications of the gravitational particle production.
All particles whose masses are smaller than that of the oscillating scalar field are produced by the gravitational
particle production if they are not Weyl-invariant. Thus it is possible that the daughter particle
itself is quite massive. 
If it is stable and heavy enough, it can serve a sizable contribution to the dark matter abundance.
Alternatively, if it is a long-lived particle such as moduli, a severe constraint on the abundance
is obtained from the observation of the big-bang nucleosynthesis.
If it is massless, on the other hand, it can contribute to the dark radiation
that is constrained by the cosmic microwave background observation.
One of the well motivated examples of such a light particle is the axion.
For example, if the theory is described solely by the standard model, 
the Peccei-Quinn sector and the Starobinsky $R^2$ inflation, 
it may produce observable amount of axion dark radiation.
A detailed study on this respect may be interesting, 
which we leave as a future work.

%%%%%%%%%%%%%%%%%%%%%%%%%%%%%%%%%%%%%%%%%%%%
\section*{Acknowledgments}
%%%%%%%%%%%%%%%%%%%%%%%%%%%%%%%%%%%%%%%%%%%%

This work was supported by the Grant-in-Aid for Scientific Research on Scientific Research A (No.26247042 [KN]),
Young Scientists B (No.26800121 [KN]) and Innovative Areas (No.26104009 [KN], No.15H05888 [KN]).
This work was supported by World Premier International Research Center Initiative (WPI Initiative), MEXT, Japan. 
The work of Y.E., R.J. and K.M. was supported in part by JSPS Research Fellowships for Young Scientists.
The work of Y.E. was also supported in part by the Program for Leading Graduate Schools, MEXT, Japan.

\appendix
%%%%%%%%%%%%%%%%%%%%%%%%%%%%%%%%%%%%%%%%%%%%
\section{Particle production rate in oscillating background} \label{sec:app}
\setcounter{equation}{0}
%%%%%%%%%%%%%%%%%%%%%%%%%%%%%%%%%%%%%%%%%%%%

We consider a real scalar field $\chi$ with time dependent mass:
\begin{align}
	S = \int d^4 x \left( -\frac{1}{2}(\partial \chi)^2 - \frac{1}{2}m_\chi^2(t) \chi^2  \right).
\end{align}
Let us estimate the production rate of $\chi$ particle.
Typically $m_\chi(t)$ is proportional to powers of other coherently oscillating scalar field $\phi(t)$
whose mass scale is $m_\phi$.
Hereafter we do not assume a specific form of $m_\chi(t)$ but only assume that it is an oscillating function with frequency of $\Omega$.

%%%%%%%%%%%%%%%%%%%%%%%%%%%%%%%%%%%%%%%%%%%%%%%%%%
\subsection{Quantization}
%%%%%%%%%%%%%%%%%%%%%%%%%%%%%%%%%%%%%%%%%%%%%%%%%%

Let us expand $\chi$ as
\begin{align}
	\chi = \int \frac{d^3k}{(2\pi)^3} \chi_{\vec k} e^{i\vec k \cdot \vec x}.
\end{align}
From reality condition $\chi^*=\chi$, we have $\chi_{\vec k}^* = \chi_{-\vec k}$.
The equation of motion of Fourier mode is given by
\begin{align}
	\ddot \chi_{\vec k} + \omega_k^2(t) \chi_{\vec k} = 0.   \label{ddotchik}
\end{align}
where $\omega_k^2 \equiv k^2 + m_\chi^2(t)$. Now we write $\chi_k$ in terms of ladder operator as
\begin{align}
	\chi_{\vec k} = a_{\vec k} v_{\vec k}(t) + a^\dagger_{-\vec k} v_{\vec k}^*(t),
\end{align}
where $v_{\vec k}(t)$ and $v_{\vec k}^*(t)$ are independent solutions of \eqref{ddotchik}.
Note that we should have $v_{\vec k}= v_{-\vec k}$ to satisfy the reality condition.
By using the freedom to choose overall normalization of $v_{\vec k}(t)$ and $v_{\vec k}^*(t)$, 
we can take $a_{\vec k}$ and $a^\dagger_{\vec k}$ so that they satisfy the following commutation relation,
\begin{align}
	\left[a_{\vec k}, a_{\vec k'}^\dagger\right] = (2\pi)^3\delta(\vec k- \vec k'), ~~~~
	\left[a_{\vec k}, a_{\vec k'}\right] = \left[a_{\vec k}^\dagger, a_{\vec k'}^\dagger\right] = 0.
\end{align}
On the other hand, we must have the following canonical commutation relation:
\begin{align}
	\left [\chi(\vec x), \dot\chi(\vec x')\right] = i \delta(\vec x - \vec x').
\end{align}
From this, we obtain 
\begin{align}
	v_{\vec k}\dot v_{\vec k}^* - v_{\vec k}^*\dot v_{\vec k} = i.  \label{vvdot}
\end{align}

Now let us assume the solution of the form
\begin{align}
	v_{\vec k}(t) = \frac{1}{\sqrt{2\omega_k}}\left[ 
		\alpha_{\vec k}(t) \,e^{-i\int_0^t dt' \omega_k(t')} + \beta_{\vec k}(t)\, e^{i\int_0^t dt' \omega_k(t')}
	\right].
\end{align}
There is a functional degree of freedom to impose arbitrary condition between $\alpha_{\vec k}(t)$ and $\beta_{\vec k}(t)$.
We choose it as
\begin{align}
	\dot\alpha_{\vec k} = \frac{\dot\omega_k}{2\omega_k} e^{2i\int_0^t dt' \omega_k(t')}\beta_{\vec k},~~~
	 \dot\beta_{\vec k}  = \frac{\dot\omega_k}{2\omega_k} e^{-2i\int_0^t dt' \omega_k(t')}\alpha_{\vec k},     \label{alphadot}
\end{align}
with $\alpha_{\vec k}(0)=1$ and $\beta_{\vec k}(0)=0$ to satisfy the initial condition
\begin{align}
	v_{\vec k}(t\to 0) \simeq \frac{1}{\sqrt{2\omega_k}}e^{-i\omega_kt},~~~
	\dot v_{\vec k}(t\to 0) \simeq -i\sqrt{\frac{\omega_k}{2}}e^{-i\omega_kt}.
\end{align}
Under these conditions, $\dot v_{\vec k}$ is expressed as
\begin{align}
	\dot v_{\vec k}(t) = -i\sqrt{\frac{\omega_k}{2} }\left[ 
		\alpha_{\vec k}(t) \,e^{-i\int_0^t dt' \omega_k(t')} - \beta_{\vec k}(t)\, e^{i\int_0^t dt' \omega_k(t')}
	\right].
\end{align}
Note that \eqref{vvdot} requires the following normalization condition
\begin{align}
	|\alpha_{\vec k}|^2 - |\beta_{\vec k}|^2 = 1,
\end{align}
which is automatically satisfied at all time once we impose the condition \eqref{alphadot}.

%%%%%%%%%%%%%%%%%%%%%%%%%%%%%%%%%%%%%%%%%%%%%%%%%%
\subsection{Production rate}
%%%%%%%%%%%%%%%%%%%%%%%%%%%%%%%%%%%%%%%%%%%%%%%%%%

The occupation number, or the phase space distribution of of $\chi$ is given by
\begin{align}
	f_\chi(k) = \frac{1}{2\omega_k}\left(  |\dot v_k|^2 + \omega_k^2|v_k|^2\right) - \frac{1}{2} = |\beta_{\vec k}|^2.
\end{align}
Thus $f_\chi(k)=0$ at $t\to 0$, but it grows after that.
The total number density is given by
\begin{align}
	n_\chi(t) = \int \frac{d^3k}{(2\pi)^3} f_\chi(k).
\end{align}

Thus the remaining task is to derive time evolution of $\beta_{\vec k}$.
It is easily calculated from \eqref{alphadot} as long as $\alpha_{\vec k} \simeq 1$ and $|\beta_{\vec k}| \ll 1$ hold.
In this case we have
\begin{align}
	\beta_{k}(t) \simeq \int_0^t dt' \frac{\dot \omega_k}{2\omega_k}e^{-2i\int_0^{t'} dt''\omega_k(t'')}
	= \int_0^t dt' \frac{m_\chi\dot m_\chi}{2\omega_k^2}e^{-2i\int_0^{t'} dt''\omega_k(t'')}
	\label{betak}
\end{align}
Recall that $m_\chi(t')$ is an oscillating function with frequency of $\Omega$.
It is not hard to imagine that time integral in \eqref{betak} cancels out if $\Omega$ and $\omega_k$ are much different from each other.
However, if $\omega_k \simeq \Omega$, the time integral gives linearly growing result with $t$.

To see only the time growing part, we perform integration by parts and assume $k^2 \gg m_\chi^2$ to rewrite \eqref{betak} as
\begin{align}
	\beta_{k}(t) \simeq \frac{i}{2\omega_k}\int _0^t dt' m_\chi^2(t')e^{-2i\omega_k t'}.
	\label{betak_BBS}
\end{align}
Now we consider a frequency range
\begin{align}
	\Omega-\Delta\Omega \lesssim \omega_k \lesssim \Omega+\Delta\Omega.
\end{align}
At $t \lesssim 1/\Delta\Omega$, the phase of $m_\chi^2(t)$ and $e^{-2i\omega_k t}$ roughly cancel with each other
and hence $\beta_k$ in this frequency range linearly grows with $t$.
After that, however, the oscillation feature forbids further growth.
Conversely, for fixed $t$, the frequency range with $\Delta\Omega\simeq 1/t$ experienced a linear growth.
Therefore we have
\begin{align}
	f_{\vec k}(t) \simeq \frac{\tilde m_\chi^4}{4\omega_k^2} t^2~~~{\rm for}~~~\Omega-\frac{1}{t} \lesssim \omega_k \lesssim \Omega+\frac{1}{t}.
	\label{fk}
\end{align}
Here $\tilde m_\chi$ stands for the amplitude of $m_\chi(t)$.
This expression is valid as long as $f_k \ll 1$, i.e., $t \lesssim 1/(q\Omega)$ with $q\equiv \tilde m_\chi^2/\Omega^2 (\ll 1)$.
The total number density linearly grows with $t$ as\footnote{
	In the case of three-point interaction as $m_\chi^2 = \mu \phi$ (hence $\Omega=m_\phi/2$), 
	we can explicitly calculate (\ref{betak_BBS})
	and find (\ref{number}) with a numerical coefficient $C=1$. 
	For the other type of interactions, $C$ slightly deviates from 1.
}
\begin{align}
	n_\chi(t) \simeq C\frac{\tilde m_\chi^4}{32\pi} t.    \label{number}
\end{align}
This expression does not refer to the parent field $\phi$.
We only assumed that $m_\chi(t)$ is an oscillating function with frequency $\Omega$.

This result is easily understood in terms of $\rho_\phi$ and $\Gamma_\phi$, if the coherent oscillation of $\phi$ is responsible for 
the oscillating $m_\chi(t)$.
Assuming that $\phi$ is canonically normalized, the perturbative decay rate of $\phi$ into $\chi$ pair is given by
(notice that $\Omega\sim m_\phi$)\footnote{
	Again, in the case of three-point interaction as $m_\chi^2 = \mu \phi$, we find that the perturbative decay rate $\phi\to\chi\chi$
	is given by (\ref{decay:pert}) with a numerical coefficient $C=1$.
	For the other type of interactions, $C$ slightly deviates from 1.
}
\begin{align}
	\Gamma_\phi \sim \frac{C}{32\pi}\frac{\tilde m_\chi^4}{\Phi^2 m_\phi} \sim \frac{C}{32\pi}\frac{q^2 m_\phi^3}{\Phi^2},  \label{decay:pert}
\end{align}
with $\Phi$ being the amplitude of $\phi$.
Since the energy density of $\phi$ is given by $\rho_\phi \simeq m_\phi^2\Phi^2/2$, we obtain
\begin{align}
	n_\chi(t) \simeq 2C\frac{\rho_\phi \Gamma_\phi}{m_\phi}t \sim C\frac{\tilde m_\chi^4}{32\pi} t.
\end{align}
%%

%%%%%%%%%%%%%%%%%%%%%%%%%%%%%%%%%%%%%%%%%%%%%%%%%%
\subsection{Gravitational production rate}
%%%%%%%%%%%%%%%%%%%%%%%%%%%%%%%%%%%%%%%%%%%%%%%%%%

Now let us consider the production of $\chi$ field which couples to $\phi$ gravitationally:
\begin{align}
	S = \int d^4x \sqrt{-g} \left( \frac{1}{2}f(\chi) R-\frac{1}{2}(\partial \chi)^2 - \frac{1}{2}m^2\chi^2 \right).
\end{align}
Here $m$ is constant and assumed to be smaller than the Hubble scale so that we can neglect it,
and $f(\chi)$ is a function of $\chi$.
In the minimal case we have $f(\chi) = M_P^2$ and hence 
$\chi$ feels the background oscillation only through the Hubble parameter or the scale factor.
By using the conformal time $d\tau = dt/a$ and defining $\tilde\chi\equiv a\chi$, 
it is rewritten as
\begin{align}
	S=\int d\tau d^3x \frac{1}{2}\left[ \tilde\chi'^2 -(\partial_i\tilde\chi)^2 +\frac{a''}{a}\left(\tilde\chi^2+6a^2f(\chi)\right)\right],
\end{align}
where we have dropped the mass term because we consider the case $m_\phi \gg m$ from now.
It is seen that $\tilde\chi$ generally obtains a mass of $\sim a''/a$ $(= a^2 R/6)$,
and it is oscillating function if there is a coherently oscillating scalar field as repeatedly shown in the main text,
which leads to $\tilde\chi$ particle production.
Note that the scale factor dependence vanishes in the conformal coupling 
$f(\chi)=-\chi^2/6 + M_P^2$.
Therefore there is no particle production in this case.

Below we consider the minimal case: $f(\chi) = M_P^2$.
Then we can apply the formula (\ref{number}) as a number density created within one Hubble time
by interpreting $m_\chi^2(\tau) = a''/a$. 
Thus
\begin{align}
	\frac{d \left[ a^3 n_\chi \right]}{d \tau} \simeq C\frac{(a''/a)^2}{32\pi}
	~~\to~~ \frac{d n_\chi}{dt} \simeq \frac{C}{32 \pi} 
	\left[ \frac{\ddot a}{a} + \left( \frac{\dot a}{a} \right)^2 \right]^2.  \label{nchi_grav}
\end{align}
Here we estimate $a''/a$ with its amplitude.
In the second similarity, we have omitted terms from the cosmic expansion.
If one can express $a$ in terms of $\phi$ as
\begin{align}
	a(t) = \langle a(t)\rangle \left[1 - \frac{c_n}{n} \frac{\phi^n - \langle \phi^n \rangle}{M_P^n} \right],
\end{align}
we have the $\chi$ number density produced in a time interval $t=H^{-1}$ as
\begin{align}
	n_\chi (t) \simeq \frac{C}{32\pi H}\left( \frac{c_n m_\phi^2 \Phi^{n}}{M_P^n} \right)^2.  \label{nchi_grav_phi}
\end{align}
%%

%%%%%%%%%%%%%%%%%%%%%%%%%%%%%%%%%%%%%%%%%%%%
\section{Adiabatic invariant in $f(R)$ theories} \label{sec:inv}
\setcounter{equation}{0}
%%%%%%%%%%%%%%%%%%%%%%%%%%%%%%%%%%%%%%%%%%%%

In Ref.~\cite{Ema:2015eqa}, we introduced an adiabatic invariant $J$ for the generalized Galileon theories.
This quantity satisfies $\dot{J} \sim \mathcal{O}(HJ)$ even 
when $H$ oscillates rapidly as $\dot{H} \sim \mathcal{O}(m_\text{eff}H)$.
In this appendix we generalize this quantity to $f(R)$ theories.

We consider the action~\eqref{eq:action_fr} in the absence of matter.
Using an auxiliary field $\phi$, this system is rewritten as 
\begin{align}
S
&= \int d^4x \sqrt{-g}\frac{M_P^2}{2}
\left[ f(\phi) + F(\phi) (R-\phi) \right].
\label{eq:action_fr_aux}
\end{align}
Using integration by parts, we have
\begin{align}
S
&= \int d^4x \; a^3
\frac{M_P^2}{2} 
\left[ f(\phi) - F(\phi) \phi - 6F(\phi)H^2 - 6\dot{F}(\phi)H \right],
\label{eq:action_fr_aux_bg}
\end{align}
which now has the form of the generalized Galileon action.
The adiabatic invariant can be derived by taking derivative with respect to $H$:
\begin{align}
J
&\equiv -\frac{1}{6M_P^2}\frac{\partial {\mathcal L}}{\partial H}
= FH + \frac{1}{2}\dot{F}.
\label{eq:J_fr}
\end{align}
Since we have $\phi = R$ from the action (\ref{eq:action_fr_aux_bg}),
$F$ in Eq.~(\ref{eq:J_fr}) is understood as $F(R)$ with $R$ being the background value $R = 12H^2 + 6\dot{H}$.

%%%%%%%%%%%%%%%%%%%%%%%%%%%%%%%%%%%%%%%%%%%%%%%%%%
\small
\bibliography{ref}
%%%%%%%%%%%%%%%%%%%%%%%%%%%%%%%%%%%%%%%%%%%%%%%%%%

\end{document}